\documentclass[iop]{emulateapj}

\newcommand{\kms}{\rm km~s^{-1}}
\newcommand{\kmsmpc}{\rm km~s^{-1}~Mpc^{-1}}
\newcommand{\dn}{D_{n}4000}

\slugcomment{\bf Last updated:\today}

\usepackage{color}
\usepackage{multirow}
\usepackage{natbib}

\begin{document}

\title{THE VELOCITY DISPERSION FUNCTION OF VERY MASSIVE GALAXY CLUSTERS: ABELL 2029 AND COMA}

\author{Jubee Sohn$^{1}$,
        Margaret J. Geller$^{1}$, H. Jabran Zahid$^{1}$, Daniel G. Fabricant$^{1}$, 
        Antonaldo Diaferio$^{2,3}$, Kenneth J. Rines$^{4}$ }

\affil{$^{1}$ Smithsonian Astrophysical Observatory, 60 Garden Street, Cambridge, MA 02138, USA}
\affil{$^{2}$ Universita’ di Torino, Dipartimento di Fisica, Torino, Italy}
\affil{$^{3}$ Istituto Nazionale di Fisica Nucleare (INFN), Sezione di Torino, Torino, Italy}
\affil{$^{4}$ Department of Physics \& Astronomy, Western Washington University, Bellingham, WA 98225, USA}

%=============================================================
\begin{abstract}
Based on an extensive redshift survey for galaxy cluster Abell 2029 and Coma, 
 we measure the luminosity functions (LFs), stellar mass functions (SMFs)
 for the entire cluster member galaxies.  
Most importantly, we measure the velocity dispersion functions (VDFs)
 for quiescent members. 
The MMT/Hectospec redshift survey for galaxies in A2029
 identifies 982 spectroscopic members; 
 for 838 members we derive the central velocity dispersion from the spectroscopy. 
Coma is the only other cluster surveyed as densely. 
The LFs, SMFs and VDFs for A2029 and Coma are essentially identical.
The SMFs of the clusters are consistent with simulations. 
%The VDFs for A2029 and Coma
% differ significantly from published VDFs of field galaxies. 
The A2029 and Coma VDFs for quiescent galaxies
 have a significantly steeper slope than those of field galaxies 
 for velocity dispersion $\lesssim 100~\kms$. 
The cluster VDFs also exceed the field at velocity dispersion $\gtrsim 250~\kms$. 
The differences between cluster and field VDFs 
 are potentially important tests of simulations and 
 of the formation of structure in the universe.
\end{abstract}
%=============================================================
\section{INTRODUCTION}

Statistical analyses of galaxy properties
 provide fundamental tests of
 structure formation models. 
In the standard $\Lambda$CDM model, 
 dark matter (DM) halos govern structure formation. 
Once DM halos form, 
 baryonic physics plays a role in forming galaxies 
 within the DM halo. 
Because all observable quantities are related to baryonic matter,
 finding connections between baryonic matter and the DM halo
 is a central goal.
In particular, identifying the best DM halo tracers among observables
 is a key issue. 
  
The luminosity of individual galaxies is a fundamental observable. 
The luminosity function has conventionally been used 
 to test structure formation models (e.g. \citealp{Klypin99, Vale04, Vale06, Yang08}). 
However, connecting 
 the luminosity function to the mass distribution of 
 DM halos is non-trivial.
 
Stellar mass has recently come into the spotlight
 as a better tracer of halo masses (e.g. \citealp{Behroozi10, More11, Li13, Tinker16}). 
Like luminosity, 
 stellar mass is also governed by baryonic physics, 
 but it appears to be more closely related to the DM properties \citep{More11, Li13}. 
Thus, the stellar mass function is 
 a basis for matching galaxies to DM subhalos
 (e.g. the abundance matching technique, \citealp{Kravtsov04, Conroy06, Guo10}) 
 when comparing observations with cosmological simulations. 
Several studies investigate the stellar mass function dependence on 
 galaxy morphology, color, redshift and environment \citep{Calvi13, Vulcani13}; 
 there appears to be little dependence on any of these parameters. 

Central velocity dispersion may be a more fundamental tracer of 
 the DM halo \citep{Wake12, vanUitert13, Bogdan15, Zahid16c}. 
The central velocity dispersion reflects the stellar kinematics
 governed by the central gravitational potential well. 
Moreover, the velocity dispersion is a direct dynamical measurement 
 whereas luminosity and stellar mass measurements suffer from 
 various systematic issues and model dependence (e.g. \citealp{Conroy09}). 

Taking advantage of huge galaxy surveys,
 many studies analyze statistical properties of galaxies in the general field 
 including luminosity functions (e.g. \citealp{Blanton01, Loveday12, McNaught14}),
 stellar mass functions (e.g. \citealp{Vulcani11, Vulcani13, Mortlock15}), and 
 velocity dispersion functions 
 (e.g. \citealp{Sheth03, Choi07, Chae10, MonteroDorta16}).
Using these large samples, 
 the effects of galaxy morphology, environment, and redshift 
 have also been investigated. 
However, computation of the luminosity, stellar mass, and velocity dispersion functions
 for an identical sample of galaxies is rare. 
\citet{Bernardi10} explore 
 the luminosity, stellar mass and velocity dispersion function simultaneously 
 based on the Sloan Digital Sky Survey (SDSS) data release 6 (DR6). 
They sample all environments covering a wide redshift range. 

Galaxy clusters offer another testbed 
 for the statistical study of galaxy properties.
Because galaxies in a cluster are 
 essentially at a fixed redshift and 
 share a common dense environment,
 samples of cluster galaxies complement samples from general surveys. 
Statistical studies of spectroscopically confirmed cluster members
 control for some observational biases.  
Nonetheless, 
 there are few studies that explore 
 the luminosity function \citep{Rines08, Agulli14, Agulli16, LeeY16} or 
 the stellar mass function \citep{Ferrarese16}
 of spectroscopically identified members. 
In contrast, 
 there are many studies based on photometrically determined membership 
 (e.g. \citealp{Barkhouse07, Moretti15, Lan16, Lee16}).
 
Measurements of the velocity dispersion function
 for an individual galaxy cluster are rare.  
Only \citet{Munari16} examine the velocity dispersion function
 in a galaxy cluster, A2142. 
They use SDSS spectra to measure the velocity dispersions. 
Then, they convert the velocity dispersions into circular velocities 
 to construct a circular velocity function.  
Finally, they compare the circular velocity function of A2142 
 to the circular velocity functions of subhalos in a set of numerical simulations.  
They take this approach
 because direct calculation of the velocity dispersion from simulations 
 to mimic the observations are not yet available. 
They suggest that 
 current numerical simulations underestimate 
 the number of massive ($> 200~\kms$) subhalos. 

Here, 
 we investigate the luminosity, stellar mass, and 
 velocity dispersion functions for Coma and A2029, two very massive clusters 
 ($> 4 \times 10^{14} M_{\odot}$, \citealp{Rines16}).
Our analysis is based on an essentially complete sample of  
 $\sim 1000$ spectroscopically identified members in each system.   
Comparisons among these observables 
 provide a basis for modeling 
 the connection between DM halos which are possibly traced by 
 the central velocity dispersion \citep{Bogdan15, Zahid16c}. 
We describe the data in Section 2
 and the member selection in Section 3. 
We investigate the luminosity, stellar mass, and
 velocity dispersion functions in Section 4. 
We discuss the results in Section 5 and summarize in Section 6. 
We adopt the standard cosmology of 
 $H_{0} = 70~\kmsmpc$, $\Omega_m = 0.3$ and $\Omega_{\Lambda} = 0.7$ throughout.

%=============================================================
\section{OBSERVATIONS OF A2029 AND COMA}

Abell 2029 ($z=0.078$) and Coma ($z=0.023$)
 are two of the most massive galaxy clusters in the nearby universe. 
Thus, they are ideal for studying  
 the luminosity, stellar mass, and velocity dispersion functions 
 for large samples of cluster members. 
Table \ref{cluster} summarizes the basic properties of the two clusters.  

\begin{deluxetable*}{lcccccccc}
\tablecolumns{9}
\tabletypesize{\scriptsize}
\tablewidth{0pt}
\setlength{\tabcolsep}{0.05in}
\tablecaption{Basic Properties of the Coma and A2029}
\tablehead{
\multirow{2}{*}{Name} & \colhead{R.A.\tablenotemark{a}} & \colhead{Decl.\tablenotemark{a}} &
\multirow{2}{*}{$z$\tablenotemark{a}}  & \colhead{$\sigma_{cl}$\tablenotemark{b}} & 
\colhead{R$_{200}$\tablenotemark{a}}   & \colhead{M$_{200}$\tablenotemark{a}} & 
\multirow{2}{*}{$N_{\rm mem}$} & \colhead{$N_{\rm mem}$} \\
                      & \colhead{(J2000)} & \colhead{(J2000)} & 
                      & \colhead{($\kms$)}      & 
\colhead{(Mpc)}       & \colhead{($10^{15}~{\rm M}_{\odot}$)} & & \colhead{($R_{cl} < R_{200}$)}}
\startdata
Coma  & 13:00:23.8 & +27:56:39 & 0.0235 & $947 \pm 31$ & $2.23^{+0.08}_{-0.09}$ & $1.29^{+0.15}_{-0.15}$ & 1224 & 856 \\
A2029 & 15:10:57.2 & +05:45:16 & 0.0784 & $973 \pm 31$ & $1.97^{+0.20}_{-0.21}$ & $0.94^{+0.30}_{-0.27}$ &  982 & 518
\enddata
\tablenotetext{a}{Estimated from the caustic technique.}
\tablenotetext{b}{Calculated using spectroscopically identified members within $R_{200}$.}
\label{cluster}
\end{deluxetable*}

\subsection{A2029}

A2029 is a massive cluster with a dominant cD galaxy (IC 1101).
The cD has an extremely large halo ($\sim 600$ kpc, \citealp{Uson91}). 
The velocity dispersion of the cD
 increases with clustercentric distance \citep{Fisher95}. 
Although A2029 appears relaxed in the optical, 
 X-ray observations reveal an extended X-ray sloshing spiral structure
 indicative of complex internal dynamics in the intracluster medium 
 \citep{Clarke04, Walker12, Paterno-Mahler13}. 
To study the statistical properties of A2029 member galaxies, 
 we carry out a redshift survey 
 using Hectospec \citep{Fabricant08},
 a wide-field spectrograph on the 6.5m MMT.

\subsubsection{Photometric Data}
We use photometric data from SDSS DR12 \citep{Alam15}
 as the basis for the redshift survey. 
We first select extended sources brighter than $r=22$ mag 
 within a projected radius $R_{cl} < 100 \arcmin$
 from the cluster center (R.A., Decl. : $227.73813$, $5.7544$).  
Unfortunately, the SDSS DR12 photometric catalog is incomplete 
 in the north-eastern part of A2029
 where there is a small patch of missing photometry.  
To correct for the incompleteness of the SDSS DR12 photometry, 
 we compiled SDSS DR7 photometry 
 for the missing region. 
 
We use composite model magnitudes
 as the SDSS webpages recommend. 
The composite model magnitude is 
 a linear combination of de Vaucouleurs and exponential magnitudes,
 yielding an approximately Petrosian magnitude. 
We apply the extinction correction for each band provided in the SDSS photometric catalog. 
Hereafter, all magnitudes denote extinction corrected composite model magnitudes. 

\begin{figure}
\centering
\includegraphics[scale=0.5]{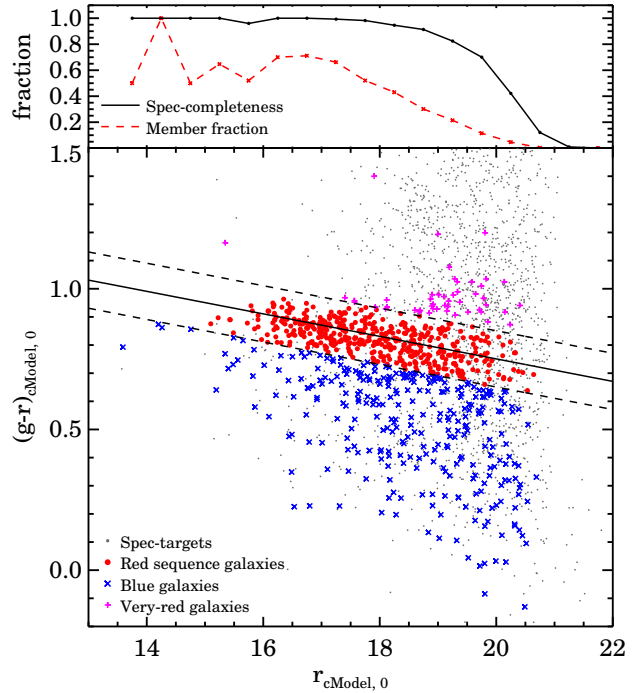}
\caption{
Color-magnitude ($g-r$ vs. $r$) diagram for A2029. 
Filled circles, crosses and pluses represent 
 red sequence, blue, and very red member galaxies, respectively. 
Gray dots show A2029 survey field galaxies with redshifts. 
The solid line shows the red sequence of A2029 and 
 the dashed lines show the boundaries of red sequence
 following \citet{Rines08}. }
\label{cmd}
\end{figure}

Figure \ref{cmd} shows the color-magnitude diagram for the A2029 region. 
We plot spectroscopic targets (see Section \ref{spec}) and 
 spectroscopically identified members (see Section \ref{msel}). 
Following \citet{Rines13}, 
 we identify the red-sequence of A2029
 by assuming a slope of $-0.04$ in the $g-r$ versus $r$ color-magnitude diagram.
We consider objects within $\pm0.1$ of the relation 
 as potential red-sequence members. 
Hereafter, we refer to the galaxies bluer/redder than the red sequence
 as blue and very red galaxies.  
 
\subsubsection{Spectroscopy}\label{spec}

Spectroscopic redshifts are the best way to determine cluster membership. 
Contamination by foreground and background objects
 is significantly reduced relative to samples based on photometric redshifts 
 (e.g. \citealp{Hwang14, Geller14}). 
Applying the caustic technique \citep{Diaferio97,Diaferio99,Serra13} 
 to the spectroscopic sample (see Section \ref{msel})
 identifies cluster members; 
 the completeness of membership determination from the caustic technique is 95\%
 within $3R_{200}$ based on numerical simulations \citep{Serra13}. 
 
For bright A2029 galaxies we compiled redshifts
 from the SDSS spectroscopic survey. 
SDSS spectra are acquired through
 $3\arcsec$ fibers for galaxies brighter than $r=17.77$. 
There are 2807 SDSS redshifts in the A2029 field ($R_{cl} < 100 \arcmin$). 
The typical measurement error for SDSS redshifts is $13~\kms$. 

We also collected redshifts from the literature. 
We compiled 40 redshifts from the NASA/IPAC Extragalactic Database (NED)
 and one redshift from the 1.5m telescope on Mt. Hopkins \citep{Sohn15}. 
\citet{Tyler13} previously conducted a redshift survey for A2029
 using the MMT/Hectospec with the 270 line mm$^{-1}$ grating.
They obtained 1164 spectra
 and identified cluster members to investigate the infrared properties of A2029 galaxies. 
We collected these spectra from the MMT archive\footnote{http://oirsa.cfa.harvard.edu/archive/search/}. 
Because SDSS DR12 revisited the A2029 area after the Hectospec survey, 
 there are 296 objects with both Hectospec and SDSS spectroscopy in this subsample. 
We use these spectra for relative calibration of the velocity dispersions
 derived from the two samples. 

We carried out our redshift survey of A2029
 using MMT/Hectospec between April and June 2016. 
The Hectospec instrument mounted on the MMT 6.5m telescope \citep{Fabricant05}
 is a multi-object fiber-fed spectrograph with 300 fibers
 covering an $\sim 1 {\rm deg}^{2}$ field of view, i.e. $R_{\rm Hecto} = 30 \arcmin$. 
We used the 270 line mm$^{-1}$ grating and the resulting spectra cover
 the wavelength range $\lambda = 3700 - 9150~{\rm \AA}$
 with a resolution of $6.2~{\rm \AA}$. 
For each Hectospec field we obtained three sequential exposures of 1200s each.
\citet{Tyler13} used the same integration times for their Hectospec observations;
 thus we include their spectra and measurements without any correction. 

We select galaxies brighter than $r=21.3$ from the SDSS photometric galaxy catalog 
 as Hectospec targets. 
We exclude galaxies with fiber magnitude $r_{\rm fib} > 22$ from the target list;
 these galaxies have a surface brightness
 too low to yield a reliable Hectospec redshift with our integration time. 
We apply no color selection to the target list.  

We reduce the data with the HSRED v2.0 package, 
 a Hectospec pipeline developed by Richard Cool. 
We measure the redshifts using RVSAO \citep{Kurtz98},
 which cross-correlates the spectra with a set of templates 
 constructed for this purpose \citep{Fabricant05}. 
We visually inspect all spectra and 
 divided the cross-correlation results into three groups: 
 `Q' for high-quality redshift, `?' for marginal cases, `X' for poor fits. 
We use only `Q' type spectra. 
In total, 
 we obtain 1597 high-quality redshifts with 
 a median measurement error of $29~\kms$.  
  
\begin{figure}
\centering
\includegraphics[scale=0.5]{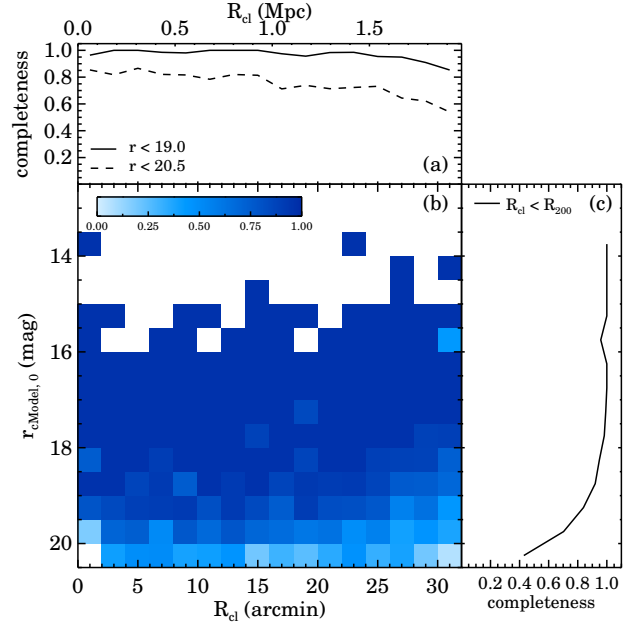}
\caption{
Two-dimensional fractional completeness of the A2029 redshift survey 
 for $R_{cl} < R_{200}$ and $r < 20.5$. 
Solid and dashed lines in the upper panels show the completeness 
 to $r=19.0$ and $r=20.5$, respectively. 
In the right panel, 
 the solid line displays the completeness 
 as a function of magnitude within $R_{cl} < R_{200}$. }
\label{zcomp}
\end{figure}

Figure \ref{zcomp} shows 
 the spectroscopic completeness of the A2029 field
 as a function of radius and $r-$band magnitude.
The spectroscopic survey is remarkably complete
 for $r < 19.0$ and $R_{cl} < 30 \arcmin$.
The spectroscopic completeness drops rapidly 
 for fainter magnitudes and for radii larger than $R_{\rm Hecto} \sim 30\arcmin$. 

\subsubsection{Stellar masses} 

To consistently compare our results with previous studies, 
 we measure stellar masses using the LePHARE
 \footnote{http://www.cfht.hawaii.edu/$\sim$arnouts/LEPHARE/lephare.html} code 
 developed by Arnout \& Ilbert \citep{Arnouts99, Ilbert06}.
We fit the stellar energy distribution (SED) measured 
 from SDSS broadband photometry 
 with stellar population synthesis (SPS) models 
 to determine the mass-to-light ratio. 
We use the mass-to-light ratio to convert the observed luminosity 
 into an estimate of stellar mass. 
We adopt the SPS models of \citet{Bruzual03} 
 and a \citet{Chabrier03} IMF.
The SPS models have two metallicities. 
We generate synthetic SEDs by varying the star formation history (SFH), 
 extinction and stellar population age. 
We adopt exponentially declining SFHs with e-folding times 
 ranging between 0 and 30 Gyr, the \citet{Calzetti00}
 extinction law with $E(B-V)$ ranging between 0 and 0.6, and 
 stellar population ages between 0.01 and 13 Gyr. 
We generate a probability distribution function (PDF) 
 for the stellar masses by $\chi^2$-fitting the synthetic SEDs 
 to the observed photometry. 
We adopt the median of the PDF as our estimate of the stellar mass.
 
Stellar masses calculated from broadband photometry 
 carry absolute uncertainties of $\sim0.3$ dex \citep{Conroy09}. 
Our application relies only on the \emph{relative} accuracy of 
 the stellar mass estimates (see \citealp{Zahid16c}). 
 
\subsubsection{Velocity dispersions}

For galaxies with SDSS spectroscopy, 
 we take velocity dispersions from the 
 Portsmouth reduction \citep{Thomas13},
 because they are consistent with 
 the velocity dispersion measured from Hectospec \citep{Fabricant13, Zahid16c}. 
\citet{Thomas13} measure the velocity dispersion 
 using the Penalized Pixel-Fitting (pPXF) code \citep{Cappellari04} 
 and the stellar population templates from \citet{Maraston11}.
These templates are 
 based on the MILES stellar library \citep{Sanchez06}. 
They convert the templates to the SDSS resolution 
 and derive the best-fit velocity dispersion. 
The median uncertainty of velocity dispersion measurement 
 from SDSS spectroscopy is $7~\kms$.

We measure velocity dispersions for all of the galaxies 
 with Hectospec spectroscopy,
 (our targets and the targets of \citet{Tyler13})
 using the University of Lyon Spectroscopic analysis Software (ULySS, \citealp{Koleva09}). 
ULySS compares the Hectospec spectra 
 with stellar population templates 
 calculated with the PEGASE-HR code \citep{LeBorgne04}
 and the MILES stellar library.
The templates are convolved to the Hectospec resolution
 at varying velocity dispersions. 
They are parameterized by age and metallicity. 
ULySS determines the best-fit age, metallicity, and velocity dispersion
 from a chi-square fit of the convolved templates to each spectrum. 
We limit the fit to the rest-frame spectral range $4100 - 5500 {\rm \AA}$.
This spectral range minimizes the uncertainty in the velocity dispersion \citep{Fabricant13}. 
The median uncertainty of the velocity dispersion measurement is $\sim 17~\kms$. 
 
Because SDSS and Hectospec spectroscopy 
 are obtained through fibers of $3\arcsec$ and $1.5\arcsec$ apertures, respectively, 
 an aperture correction is necessary \citep{Zahid16c}:
\begin{equation}
 (\sigma_{\rm SDSS} / \sigma_{\rm Hecto}) = (R_{\rm SDSS} / R_{\rm Hecto})^{\beta}
\end{equation}
Following \citet{Zahid16c}, 
 we determine the aperture correction using the 169 objects 
 with both SDSS and Hectospec velocity dispersions
 in the range $100 < \sigma < 450~\kms$ and with a velocity dispersion uncertainty smaller than $100~\kms$.
The coefficient for the aperture correction is $\beta = -0.054 \pm 0.005$,
 consistent with previous determinations: 
 $\beta = -0.066 \pm 0.035$ from \citet{Cappellari04} and 
 $\beta = -0.046 \pm 0.013$ from \citet{Zahid16c}.
This correction is small; 
 our results are not sensitive to the value of $\beta$.  
 
Below we compare the velocity dispersion functions of A2029 and Coma. 
These two target clusters are located 
 at different redshifts. 
Thus the velocity dispersions through the 
 fiber apertures trace different portion of the target galaxies. 
Therefore, we correct the velocity dispersions to 
 a fiducial physical aperture of 3 kpc following \citet{Zahid16c}. 
We employ equation (1) again for this process. 
Hereafter, all central velocity dispersion, $\sigma$, 
 represents the value within a 3 kpc (rest-frame) aperture. 

\subsubsection{$\dn$} 

The $\dn$ index is defined as the ratio of flux in two spectral windows 
 adjacent to the 4000${\rm \AA}$ break \citep{Balogh99}.
We calculate the index by taking the flux (measured per unit frequency) 
 in the interval 4000-4100${\rm \AA}$ relative to the flux in the interval 3850-3950${\rm \AA}$. 
The $\dn$ index is sensitive to the stellar population age \citep{Kauffmann03, Geller14}. 
The $\dn$ index also has some metallicity dependence \citep{Kauffmann03, Woods10}.
Because the distribution of $\dn$ is bimodal, 
 $\dn$ can be used to separate star-forming and quiescent galaxies spectroscopically
 \citep{Mignoli05, Vergani08, Woods10}. 

For SDSS galaxies, 
 we adopt the $\dn$ value from the MPA/JHU catalog\footnote{http://www.mpa-garching.mpg.de/SDSS/DR7/}. 
For BOSS and Hectospec data we calculate $\dn$ directly from the spectra. 
\citet{Fabricant08} %Fabricant et al. (2008, PASP, 120, 1222) 
 show that $\dn$ measured from Hectospec and SDSS spectroscopy are consistent to within $\sim5\%$. 
This level of consistency is sufficient for our application.

\subsection{Coma}

We also use SDSS DR12 photometry for Coma galaxies. 
Because of the proximity of Coma, 
 we construct the galaxy sample from a larger area $R_{cl} < 300 \arcmin$,
 corresponding to $100 \arcmin$ for A2029.  
As in the A2029 sample,
 we use extinction-corrected composite model magnitudes.

\begin{figure}
\centering
\includegraphics[scale=0.5]{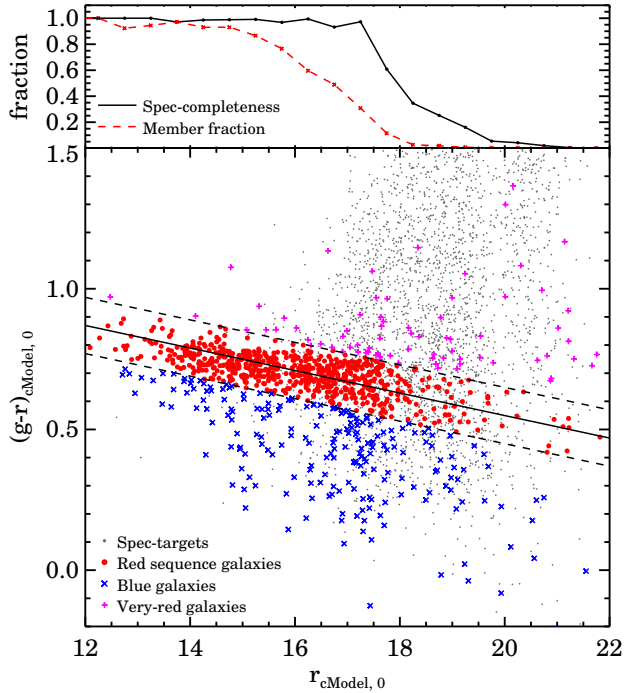}
\caption{Same as Figure \ref{cmd}, but for Coma. }
\label{cm_cmd}
\end{figure}

Figure \ref{cm_cmd} displays 
 the color-magnitude diagram for Coma. 
We determine the red-sequence using the same method applied to A2029. 
The red-sequence appears bluer than for A2029 
 because Coma is three times closer. 
The red-sequence of the two clusters appears in the same color range
 if we applying appropriate $K-$correction.

\begin{figure}
\centering
\includegraphics[scale=0.5]{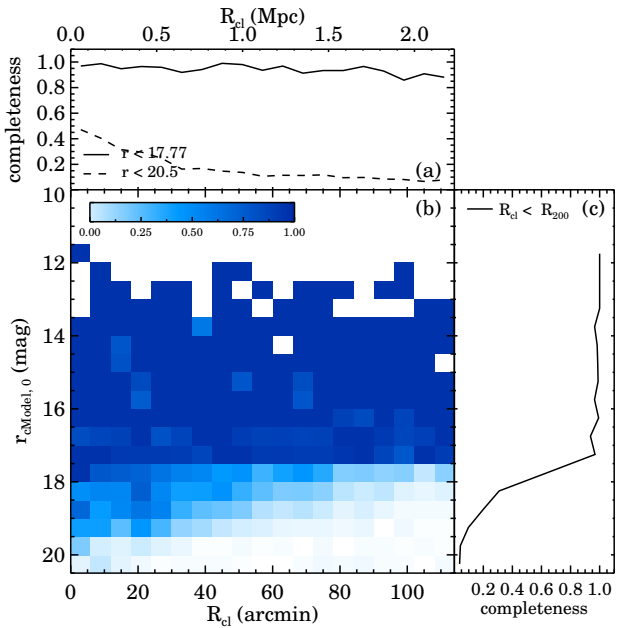}
\caption{
Same as Figure \ref{cmd}, but for Coma.
Solid line in the upper panel shows the completeness to $r=17.77$ 
 the SDSS spectroscopic survey limit. }
\label{cm_zcomp}
\end{figure}

Most redshifts for Coma come from SDSS DR12 and BOSS. 
Although SDSS surveys a huge field, 
 there is some residual incompleteness in dense regions
 due to fiber collisions \citep{Strauss02, Park09, Shen16}. 
We thus compile missing redshifts from the literature, 
 mainly from NED (for more details, see \citealp{Hwang10}). 
In total, we compile 22410 redshifts within $R_{cl} < 300 \arcmin$. 
Figure \ref{cm_zcomp} shows the spectroscopic completeness of Coma. 
The spectroscopy for the Coma field is highly complete to $r \sim 17.77$, 
 the spectroscopic limit of the SDSS. 

Using the same technique as for A2029, 
 we measure the stellar masses of Coma galaxies. 
Similar to the A2029 galaxies with SDSS spectroscopy, 
 we calculate fiducial central velocity dispersions within 3 kpc
 by applying equation (1) to the measurement of \citet{Thomas13}.
The median uncertainty of the velocity dispersions for Coma members 
 is $\sim 4~\kms$. 
$\dn$s of Coma galaxies are from the MPA/JHU catalog. 

%=============================================================
\section{CLUSTER MEMBERSHIP}\label{msel}

We analyze the distribution of galaxies in phase space to identify cluster members.
We use the caustic technique \citep{Diaferio97, Diaferio99, Serra13} to analyze the data. 
Simulations by \citet{Serra13} show that 
 the caustic technique recovers 95\% of the clusters members within $3R_{200}$ 
 for mock catalogs with $\sim 1000$ galaxies in the field of view and
 $\sim 180$ members per cluster; these objects all appear within the caustics. 
The interloper contamination from non-member galaxies that
 appear within the caustics is $\sim 2\%$ at $R_{200}$ and $\sim 8\%$ at $3R_{200}$. 
The performance of the caustic technique may be even better for very rich clusters like 
 Coma and A2029 that contain $\sim 1000$ members. 
 
\subsection{A2029}

\begin{figure}
\centering
\includegraphics[scale=0.5]{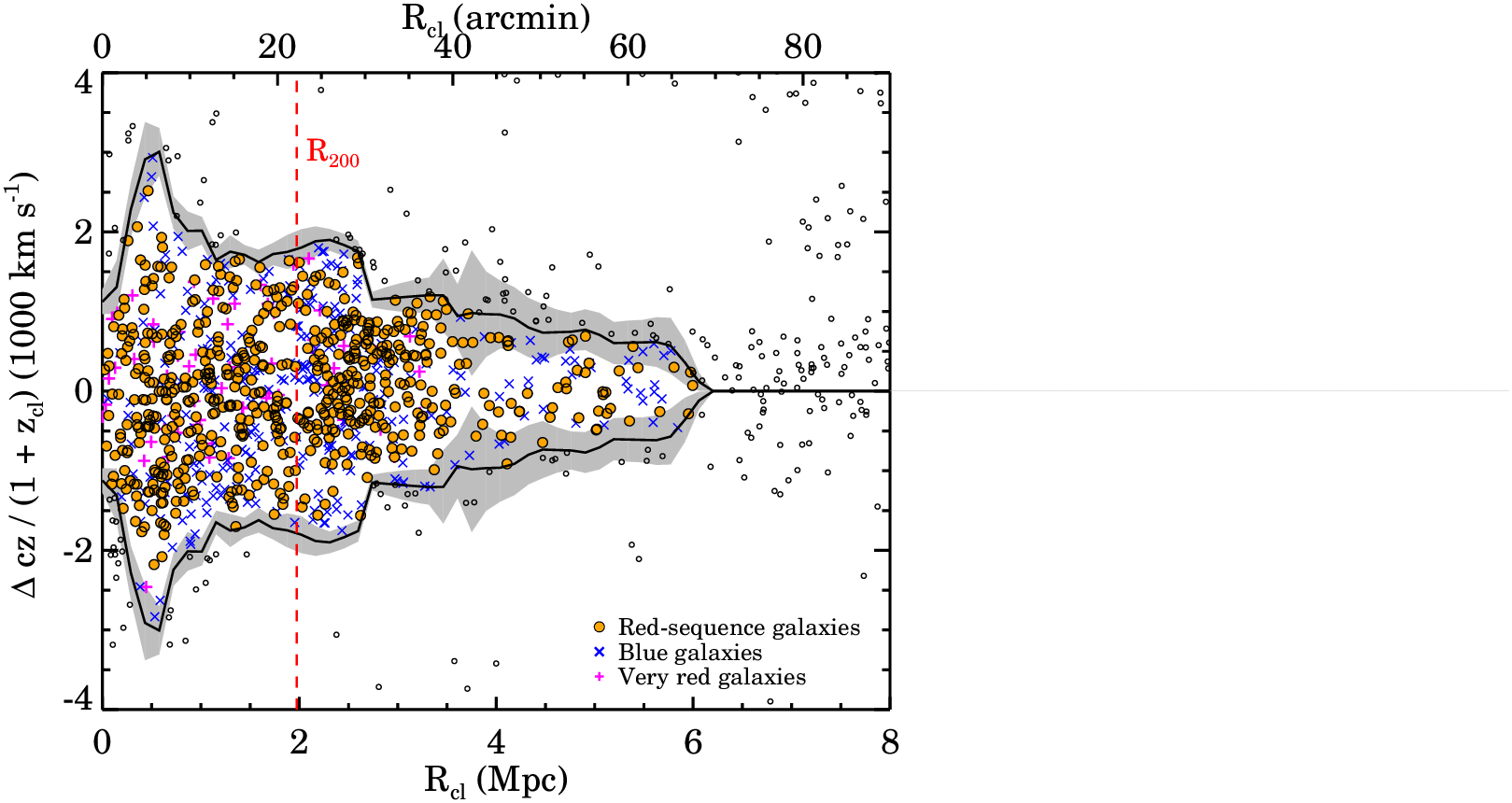}
\caption{Rest-frame clustercentric radial velocity vs. 
projected clustercentric distance for A2029. 
Filled circles, crosses and pluses represent 
 red sequence, blue and very red A2029 members, respectively. 
Open circles indicate A2029 non-members. 
The solid line shows the caustics for A2029 and 
 the shaded regions shows the uncertainty in the caustic estimate. 
The vertical dashed line shows $R_{200}$.}
\label{rv}
\end{figure}

Figure \ref{rv} shows rest-frame clustercentric velocity versus 
 projected clustercentric distance, the R-v diagram, for A2029. 
There is a clear concentration around the center of the cluster.
The non-parametric caustic technique identifies the sharp boundaries of the cluster in Figure \ref{rv}. 
We identify objects within the caustics as cluster members.
Within the caustic, 
 there are 982 members with a mean redshift $z = 0.078$.  
Figure \ref{cmd} shows that 
 60\% of cluster members are on the red sequence; 
 34\% and 6\% of the galaxies are blue and very red, respectively. 
Not all objects with red sequence colors are cluster members;  
 41\% of the spectroscopic targets on the red sequence are non-members.   
Table \ref{member} lists 
 the redshift and $\sigma$ for each A2029 member galaxy.  

\citet{Diaferio97} and \citet{Diaferio99}
 identify the caustics with the escape velocity from the cluster.
This identification in turn provides the mass profile 
 as a function of projected distance from the cluster center \citep{Serra11}. 

\begin{deluxetable*}{ccccccc}
\tablecolumns{7}
\tabletypesize{\footnotesize}
\tablewidth{0pt}
\setlength{\tabcolsep}{0.05in}
\tablecaption{A2029 Spectroscopic Members}
\tablehead{
\colhead{R.A.} & \colhead{Decl.} & \colhead{$cz$} & \colhead{$cz$ error} & 
\colhead{$\sigma\tablenotemark{a, b}$} & \colhead{$\sigma$ error} & \colhead{Ref\tablenotemark{b}} \\
\colhead{(J2000)} & \colhead{(J2000)} & \colhead{($\kms$)} & \colhead{($\kms$)} &
\colhead{($\kms$)} & \colhead{($\kms$)} & }
\startdata
227.740259 &  5.766147 & 23177 &   29 & 284 &   5 &  T13 \\
227.737793 &  5.762320 & 22763 &   22 &  95 &  17 &  T13 \\
227.744635 &  5.770809 & 22247 &    4 & 320 &   5 & SDSS \\
227.738249 &  5.754465 & 23162 &   26 & 229 &  10 &  T13 \\
227.749463 &  5.769346 & 23468 &   22 &  61 &  14 &  T13 \\
227.735039 &  5.751555 & 23797 &   35 & 261 &  42 &  MMT \\
227.732491 &  5.765348 & 23679 &   18 & -99 & -99 &  T13 \\
227.750338 &  5.782786 & 23464 &   24 & 274 &   5 &  T13 \\
227.735860 &  5.775843 & 24381 &   24 & 151 &   5 &  T13 \\
227.728122 &  5.756973 & 24019 &   26 & 160 &   7 &  T13 
\enddata
\label{member}
\tablenotetext{a}{The central velocity dispersion within a rest-frame 3 kpc aperture.}
\tablenotetext{b}{-99 indicates lack of a measurement for $\sigma$} 
\tablenotetext{c}{The redshift source:
 `SDSS' from SDSS and BOSS, `T13' from \citet{Tyler13} and 
 `MMT' from the MMT/Hectospec observation for this study. }
\tablecomments{
The entire table is available in machine-readable form in the online journal. 
Here, a portion is shown for guidance regarding its format. }
\end{deluxetable*}

From the caustics in Figure \ref{rv}, 
 we compute the characteristic mass $M_{200}$ and radius $R_{200}$
 where the mean density is 200 times the critical density of the universe. 
Table \ref{cluster} lists the measured values for A2029; 
 $R_{200} = 1.97^{+0.20}_{-0.21}$ Mpc and 
 $M_{200} = 0.94^{+0.30}_{-0.27} \times 10^{15} M_{\odot}$. 
The derived $R_{200}$ and $M_{200}$ 
 are consistent with estimates from X-ray observations;
 $R_{200} = 1.92^{+0.11}_{-0.13}$ Mpc and 
 $M_{200} = 8.0^{+1.5}_{-1.5} \times 10^{14} M_{\odot}$ \citep{Walker12}. 
The velocity dispersion within $R_{200}$, $\sigma_{cl} = 973 \pm 32~\kms$, 
 is also consistent with 
 $\sigma_{cl} = 954 ^{+76}_{-61}~\kms$ from \citet{Rines16}.
 
\subsection{Coma}

\begin{figure}
\centering
\includegraphics[scale=0.5]{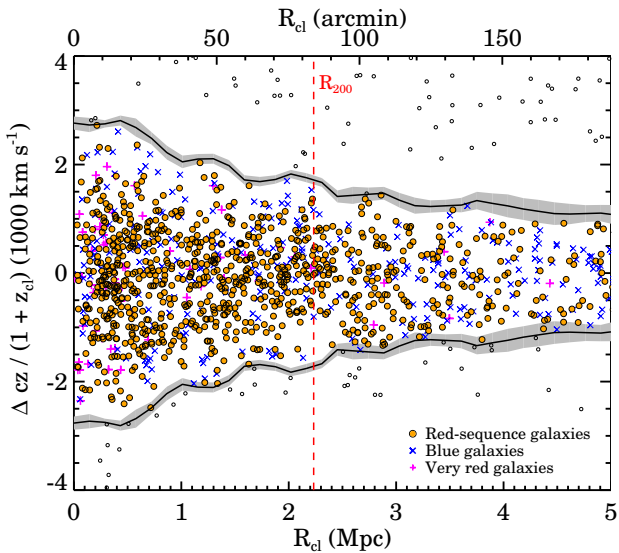}
\caption{Rest-frame clustercentric radial velocities vs. 
projected clustercentric distances for Coma.
Symbols and lines are the same as in Figure \ref{rv}. }
\label{cm_rv}
\end{figure}

Figure \ref{cm_rv} shows the R-v diagram for Coma.
The caustics of Coma look smoother than those of A2029. 
Coma galaxies extend over a very wide region
 even beyond $R_{cl} = 3$ deg. 
We restrict our plot to $R_{cl} \sim 3$ deg.
This range is still much larger than the characteristic scale of Coma
 (e.g. $R_{200} \sim 1.8^{\circ} \sim 2.1$ Mpc, \citealp{Geller99}).
Within the caustic, 
 we identify 1251 members with a mean redshift $z=0.0231$. 
The fraction of each population in Coma (Figure \ref{cm_cmd}) 
 is similar to that of A2029; 
 66\% are red-sequence galaxies, 25\% and 9\% are blue and very red members, respectively. 
  
We measure $R_{200}$ and $M_{200}$ for Coma (Table \ref{cluster}) 
 based on the caustics. 
Coma is slightly larger and more massive than A2029 with 
 $R_{200} = 2.23^{+0.08}_{-0.08}$ Mpc and 
 $M_{200} = 1.29^{+0.15}_{-0.15} \times 10^{15} M_{\odot}$. 
The $R_{200}$ is consistent with the previous caustic measurement \citep{Geller99}. 
The caustic mass is consistent with
 the weak lensing mass measurement, 
 $M_{200} = 8.9^{+3.6}_{-2.0} \times 10^{14} M_{\odot}$ \citep{Okabe14}. 
The velocity dispersion of Coma within $R_{200}$ 
 is $\sigma_{cl} = 947 \pm 31~\kms$, 
 also similar to the previous measurements:
 $\sigma_{cl} = 1082 \pm 74~\kms$ \citep{Colless96} and 
 $\sigma_{cl} = 957 ^{+30}_{-28}~\kms$ \citep{Rines03}. 

\section{LUMINOSITY, STELLAR MASS AND VELOCITY DISPERSION FUNCTION OF CLUSTER GALAXIES}

We construct the luminosity, stellar mass, and 
 central velocity dispersion functions 
 for A2029 and Coma. 
These three functions are powerful probes of the 
 mass distribution of DM subhalos, 
 crucial for modeling galaxy formation and evolution. 
Conventionally, 
 luminosity and stellar mass functions
 have been favored because they can be derived from photometric data alone. 
However, contamination by interlopers 
 can be a serious issue (e.g. \citealp{Geller14}).
Here, 
 we measure the luminosity (Section \ref{lf}) and stellar mass functions (Section \ref{smf}) 
 based on samples of spectroscopically identified members. 
We also compute the velocity dispersion functions
 for quiescent galaxies (Section \ref{vdf})
 that may provide a more direct connection to the DM halo masses
 (see \citealp{Zahid16c} and the reference therein).

\subsection{The Spectroscopic luminosity function}\label{lf}

At the faint limit, 
 the redshift surveys are not complete (see Figure \ref{zcomp} and Figure \ref{cm_zcomp}).
Thus, we must correct for spectroscopic incompleteness 
 \citep{Rines08, Agulli14}.
We follow the method of \citet{Rines08} 
 who correct for missing members
 as a function of apparent magnitude, projected distance from the cluster center, 
 and galaxy color.
 
\begin{figure}
\centering
\includegraphics[scale=0.5]{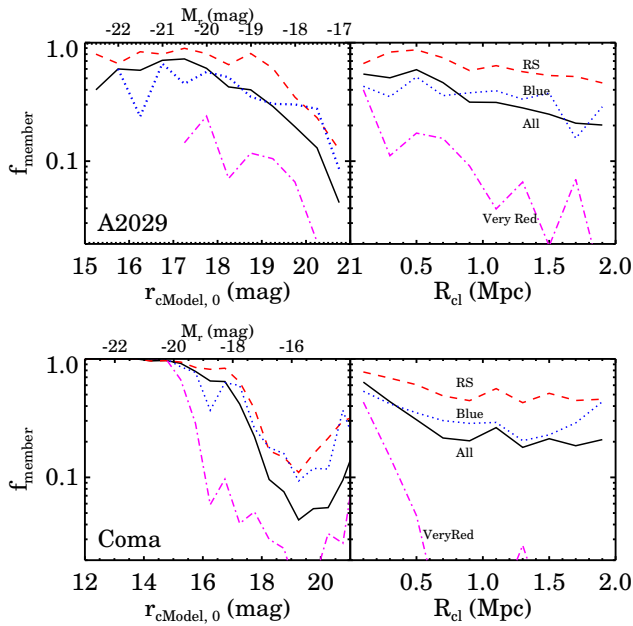}
\caption{(Upper panels) Spectroscopic member fractions for A2029 
 as a function of magnitude (left) and clustercentric distance (right). 
Lines represent the fractions for red-sequence galaxies (dashed), 
 blue (dotted) and very red (dot-dashed) galaxies. 
Solid lines show the sum of the three populations. 
(Lower panels) Same as the upper panels, but for Coma. 
Note that the behavior of three populations is similar for the two clusters. }
\label{lfcor}
\end{figure} 
 
The upper panels of Figure \ref{lfcor} show the member fractions 
 as a function of magnitude (left panel) and 
 projected clustercentric distance (right panel) for A2029. 
The member fraction is $f_{\rm member} = N_{\rm member} / N_{\rm spec}$, 
 where $N_{\rm member}$ is the number of caustic members, and
 $N_{\rm spec}$ is the number of spectroscopic targets.
We investigate the member fraction trends 
 for red sequence, blue, and very red galaxies, separately
 (see Figure \ref{cmd} for classifications).
The member fractions decline as a function of magnitude
 for all three populations (the upper left of Figure \ref{lfcor}). 
However,  
 the member fractions of the three populations behave differently, 
 emphasizing the need for separate corrections. 
The member fractions also decrease with radius. 
Again, the member fractions for the three populations differ. 
The member fraction for the very red population 
 drops more rapidly than the other populations. 
The lower panels of Figure \ref{lfcor} 
 plot the same quantities for Coma. 
The member fractions for Coma behave in essentially the same way as those for A2029. 
 
Using the member fractions in Figure \ref{lfcor}, 
 we apply corrections to account for missing members. 
We first count the number of photometric galaxies ($N_{\rm phot}$) in the three populations 
 within 0.5 magnitude bins. 
Then, we derive the corrected luminosity functions:
\begin{equation}
\phi(m_{r}) = N_{\rm phot} (m_{r}) \times \frac{N_{\rm member} (m_{r})}{N_{\rm spec} (m_{r})} \times \frac{1}{A}, 
\end{equation}
 where A is the area.
The total luminosity function is the sum of the luminosity functions for the three populations.  
We restrict our analysis to $R_{cl} < R_{200}$
 where the corrections are relatively small for both clusters. 
 
\begin{figure*}
\centering
\includegraphics[scale=0.50]{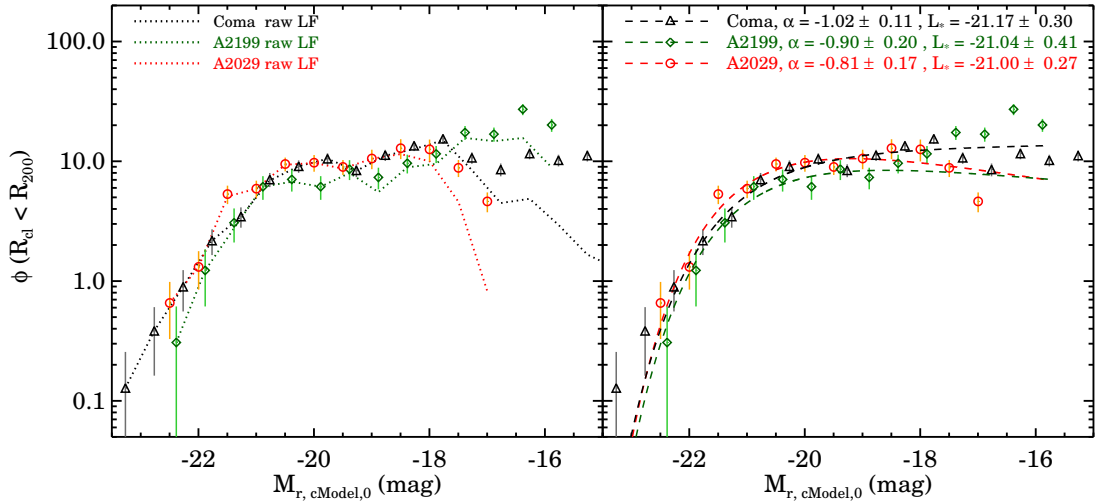}
\caption{
Spectroscopic luminosity function of 
 A2029 (circles), Coma (triangles), respectively, 
 corrected for incompleteness. 
For reference, 
 the diamonds show the LF for A2199 \citep{Rines08}. 
The left panels show the raw luminosity functions (dotted lines).  
The right panels display the 
 Schechter function fits for the magnitude range $-22 < M_{\rm r, cModel, 0} < -18$.}
\label{lfcomp}
\end{figure*} 
 
Figure \ref{lfcomp} shows 
 the $r-$band spectroscopic LFs of A2029 and Coma. 
We compare these results with the spectroscopic LF of the A2199 cluster \citep{Rines08}, 
 another massive cluster in the local universe ($z=0.03,~\sigma_{cl}=676~\kms$). 
Using the A2199 data from \citet{Rines08}, 
 we also measure the LF within $R_{cl} < R_{200}$. 
We plot the A2199 LF 
 as a sum of the three cluster populations. 

The LFs of all three clusters 
 are remarkably similar for $M_{r} < -18$
 except for small discrepancies at the very bright end. 
Coma has a few more bright galaxies at $M_{r} < -22$;
 these objects are the brightest cluster galaxies 
 within each substructure of Coma \citep{Colless96, Okabe14}. 
Thus, the increment in the Coma LF at the bright end
 probably results from the complex nature of the cluster. 
In contrast, A2029 and A2199 have a single brightest cluster galaxy. 

Towards the faint end, 
 the three LFs appear to have different shapes. 
The Coma and A2199 LFs show a slight upturn for $M_{r} \gtrsim -18.0$; 
 the A2029 LF appears to decline. 
The Coma LF shows an even stronger upturn at the faint end 
 when low surface brightness galaxies, ultra compact dwarfs, and globular clusters 
 are taken into account \citep{Milne07}. 

\begin{figure}
\centering
\includegraphics[scale=0.5]{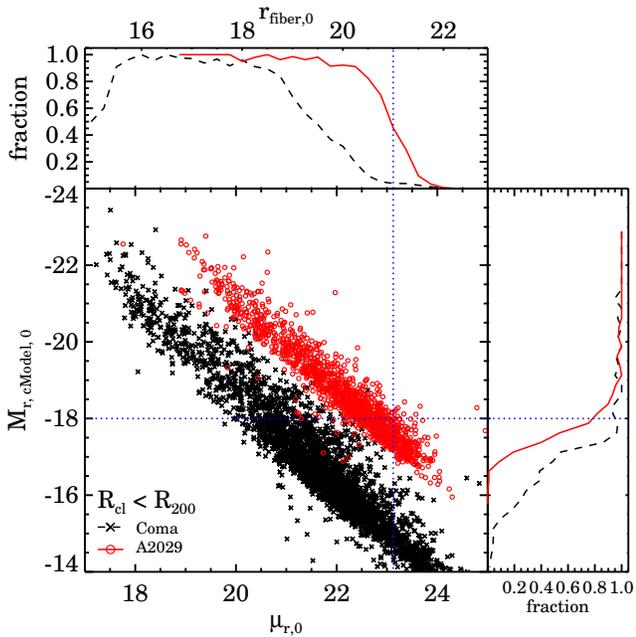}
\caption{
Absolute magnitude vs. surface brightness for 
 the spectroscopic samples in the Coma (crosses) and A2029 fields (circles).
The upper panel shows
 the spectroscopic completeness as a function of surface brightness. 
Dashed and solid lines show the completeness for Coma and A2029, respectively. 
The right panel displays the completeness but as a function of absolute magnitude.  }
\label{sbri}
\end{figure} 

The A2029 LF appears to decline at the faint end 
 as a result of the surface brightness limit of the spectroscopic sample. 
A2029 is more distant than both Coma and A2199. 
Consequently, 
 the observed surface brightness at the same absolute magnitude 
 is significantly fainter in A2029.  
Figure \ref{sbri} demonstrates this point. 
The spectroscopic completeness for A2029 
 declines rapidly at M$_{r} \sim -18$ 
 because the redshift survey is incomplete at $r_{\rm fib} > 21$.

We fit a Schechter function \citep{Schechter76}, 
\begin{equation}
\phi(M) = \phi_{0}~10^{0.4(1+\alpha)(M_{*}-M)}~{\rm exp}[-10^{0.4(M_{*}-M)}], 
\end{equation}
 to the LFs in the range $-22.0 < M_{r} < -18.0$. 
Table \ref{fit} summarizes 
 the Schechter function fitting parameters for 
 Coma, A2199 and A2029. 
The A2029 spectroscopic LF has 
% fits with the shallow Schechter function with 
 $\alpha = -0.81 \pm 0.17$ and $M_{r,*} = -21.00 \pm 0.27$.
The LFs of the other two clusters are also well described
 by Schechter functions:
 $\alpha = -1.02 \pm 0.11$ and $M_{r,*} = -21.17 \pm 0.30$ for Coma and 
 $\alpha = -0.90 \pm 0.29$ for $M_{r,*} = -21.04 \pm 0.41$ for A2199, respectively.
These slopes are consistent with the slope for 
 the spectroscopic LF of A85 ($z=0.055$, \citealt{Agulli14})
 with $\alpha = -0.67 \pm 0.25$ at the bright end $-22 < M_{r} < -19$.

\begin{deluxetable*}{lcccc}
\tablecolumns{5}
\tabletypesize{\scriptsize}
\tablewidth{0pt}
\setlength{\tabcolsep}{0.05in}
\tablecaption{The Spectroscopic Luminosity Function Parameters}
\tablehead{
\colhead{Cluster} & \colhead{Fitting range} & \colhead{$\alpha$} & \colhead{M$_{*}$} & \colhead{Ref.}}
\startdata
A2029 & $-22.0 \leq $M$_{r} \leq -18.0$ & $-0.81 \pm 0.17$ & $-21.00 \pm 0.27$ & This study \\
Coma  & $-22.0 \leq $M$_{r} \leq -18.0$ & $-1.02 \pm 0.11$ & $-21.17 \pm 0.30$ & This study \\
A2199 & $-22.0 \leq $M$_{r} \leq -18.0$ & $-0.90 \pm 0.29$ & $-21.04 \pm 0.41$ & This study \\
\hline
A2199 & $-22.5 \leq $M$_{r} \leq -16.0$ & $-1.13 \pm 0.07$ & $-21.11 \pm 0.25$ & \citet{Rines08} \\
Virgo & $-22.0 \leq $M$_{r} \leq -16.5$ & $-1.28 \pm 0.06$ & $-21.32$          & \citet{Rines08} \\
A85   & $-22.5 \leq $M$_{r} \leq -19.0$ & $-0.79 \pm 0.09$ & $-20.85 \pm 0.14$ & \citet{Agulli14} 
\enddata
\label{fit}
\end{deluxetable*}

The shape of a spectroscopic LF may
 differ from a photometrically determined LF for two reasons: 
 (1) interloper contamination tends to be greater for photometrically selected samples and 
 (2) the fitting range for the LF may not be restricted to the bright end. 
\citet{Lagana11} present photometric LFs of these three clusters
 based on SDSS $i-$band photometry. 
They fit these LFs with double Schechter functions 
 at the bright and faint ends separately. 
Their fits for the bright end (M$_{i} < -18$) are generally steeper than ours:
 $\alpha = -1.26 \pm 0.03$ for Coma, 
 $\alpha = -1.18 \pm 0.03$ for A2199, and 
 $\alpha = -1.17 \pm 0.07$ for A2029.
The difference may result from interlopers. 
We note again that 
 $> 40\%$ of the cluster members on red sequence with (M$_{i} < -18$) are 
 non-members. 
The photometric LFs include some of these interlopers and 
 are thus steeper. 
The interloper fraction increases for less luminous objects.
The photometric LFs of Coma derived by \citet{Andreon02} and \citet{Milne07} 
 are substantially steeper, 
 but the difference here is dominated by 
 their broader fitting range, M$_{R} < -9.5$.

Steeper spectroscopic LFs for Coma and A2199 in the literature
 may also result from their broader fitting range:
 $\alpha \sim -1.2$ for Coma for M$_{R} < -16$ \citep{Mobasher03}, 
 $\alpha \sim -1.1$ for A2199 for M$_{r} < -16$ \citep{Andreon08, Rines08}.   
In fact, we derive a similar slope when we fit the A2199 LF to M$_{r} < -16$. 
The slope dependence on the fitting range is well known
 \citep{Agulli14, Moretti15, Lan16}. 
\citet{Lan16} also show that cluster LFs become steeper 
 faintward of M$_{r} > -18$. 
This behavior is evident in Figure \ref{lfcomp}. 

\subsection{Stellar Mass Function}\label{smf}

The stellar mass function (SMF) appears to be a 
 more fundamental tracer of DM halo masses than the LF
 (\citealp{More11, Li13}). 
However, SMFs are less frequently measured 
 because derivation of the stellar mass requires multi-band photometry. 
Furthermore, stellar mass is not directly observable. 
The derived stellar mass is sensitive to 
 the stellar IMF and star formation history \citep{Conroy09}. 
Thus, comparison of SMFs for different clusters
 must be based on consistent stellar mass computations. 

To compute complete cluster SMFs,
 we correct for stellar masses of two types of missing members.
First, stellar mass estimation miscarries for 
 5\% of the members in A2029 and 2\% of the members in Coma, respectively. 
Second, there are missing members resulting from spectroscopic incompleteness
 (see Figure \ref{lfcor}). 

We empirically estimate 
 the stellar masses ($M_{*}$) for both types of missing members 
 based on the conditional probability distributions $P(M_{*}|M_{r})$ of 
 $M_{*}$ given the absolute magnitude for the three-different galaxy populations. 
The conditional probability distribution $P(M_{*}|M_{r})$ depends on galaxy type. 
Thus, we derive these distributions empirically for 
 red-sequence, blue and very red galaxies 
 based on $k-$corrected colors for galaxies
 in the wide-field SDSS for the redshift range $0.01 < z < 0.09$ 
 where the upper redshift limit is the A2029 redshift. 
We identify the red-sequence in the $(g-r) - r$ color magnitude diagram.
  
\begin{figure}
\centering
\includegraphics[scale=0.5]{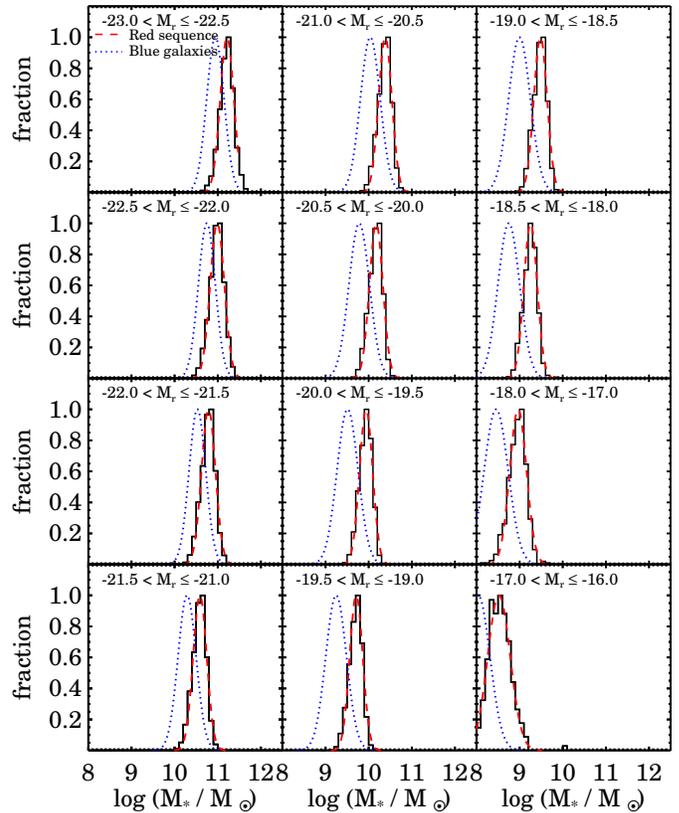}
\caption{
Conditional probability $P_{RS}(M_{*}|M_{r})$
 histograms for red sequence galaxies derived from the SDSS field sample.
The dashed and dotted lines show Gaussian fits for the probability distributions
 for red sequence and for blue galaxies. 
These distributions demonstrate the necessity of correcting 
 separately for the different populations. } 
\label{mfcor}
\end{figure}

Figure \ref{mfcor} shows the conditional probability distributions 
 $P_{rs}(M_{*}|M_{r})$ for SDSS red sequence galaxies 
 in different fixed magnitude ranges. 
The $M_{*}$ distribution for each magnitude bin 
 is well represented by a Gaussian. 
The dashed lines in Figure \ref{mfcor} show 
 the best-fit Gaussian distributions for the set of $P_{rs}(M_{*}|M_{r})$ distributions. 
The distributions, $P_{blue}(M_{*}|M_{r})$ and $P_{vr}(M_{*}|M_{r})$ 
 for SDSS blue and very red galaxies, respectively, 
 can also be described by Gaussians. 
However,
 the means and widths of the Gaussian differ for different populations. 
For example, the dotted blue lines in Figure \ref{mfcor} show
 the best-fit Gaussian distributions for the set of $P_{blue}(M_{*}|M_{r})$. 
This comparison clearly shows
 the necessity of an empirical correction that depends on both color and magnitude. 

Using the SDSS field samples for the three populations 
 covering the relevant magnitude range, 
 we compute the entire set of distributions, $P_{pop}(M_{*}|M_{r})$. 
For members where the $M_{*}$ computation failed, 
 we know the absolute magnitude. 
For members missing as a result of spectroscopic incompleteness, 
 we randomly select a galaxy in the appropriate apparent magnitude bin and then, 
 because we take the object as a cluster member, we automatically have the absolute magnitude.
In both cases we derive an $M_{*}$ for the galaxy 
 by drawing randomly from the appropriate $P(M_{*}|M_{r})$.  
We repeat the process for the entire sample of missing members 1000 times. 
Each of these 1000 sample provides an estimate of the SMF. 
We then take the mean of these 1000 SMFs 
 as the `corrected' SMF of the cluster.  
 
\begin{figure}
\centering
\includegraphics[scale=0.53]{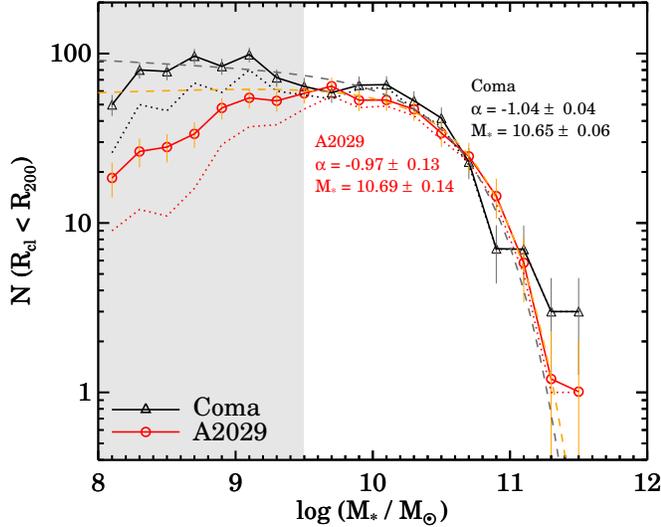}
\caption{
Stellar mass function for Coma (triangles) and A2029 (circles). 
Error bars are Poisson. 
The dotted lines show the raw mass function and 
 the dashed lines represent the Schechter function fits for each cluster. } 
\label{mf}
\end{figure}

Figure \ref{mf} displays
 the SMFs of Coma and A2029 for $R_{cl} < R_{200}$. 
The dotted and solid lines show 
 the raw and corrected SMFs, respectively. 
The corrections are significant only for $\log (M/M_{\odot}) < 9.5$
 corresponding to $M_{r} < -18.5$, 
 where the spectroscopic completeness declines. 

The two SMFs are similar for 
 $9.5 \leq \log (M_{*}/M_{\odot}) \leq 11.5$
 where the corrections for incompleteness are negligible. 
As for the LF,
 the A2029 SMF appears to decline rapidly at low $M_{*}$,
 but the Coma SMF remains flat. 
This difference is an artifact resulting from 
 the relatively lower spectroscopic incompleteness of the A2029 sample
 at faint magnitudes. 

We fit the SMFs
 with the Schechter form in the range 
 $9.5 \leq \log (M_{*}/M_{\odot}) \leq 11.5$. 
The results for Coma and A2029 MFs are; 
 $\alpha = -1.04 \pm 0.04$, $M_{*} = 10.65 \pm 0.06$ for Coma and
 $\alpha = -0.97 \pm 0.13$, $M_{*} = 10.69 \pm 0.14$ for A2029. 
The two SMFs are consistent.  
The MFs of these clusters
 are somewhat steeper than their LFs, but 
 the $M_{*}$ values are consistent with a direct conversion of the luminosity into $M_{*}$
 using the relation from the SDSS field galaxies on the red-sequence: 
 $M_{*,converted} \sim 10.52$ for Coma and 
 $M_{*,converted} \sim 10.48$ for A2029. 
 
Figure \ref{mfcor} demonstrates that 
 overall construction of the SMF by converting mean luminosities to 
 mean $M_{*}$ is inadequate. 
In a fixed magnitude range, 
 the $M_{*}$ distribution depends on galaxy type or color. 
Furthermore, 
 the $M_{*}$ range varies with the range of absolute magnitude.  
At the bright end, 
 the LF is dominated by the red sequence population 
 which has a narrow $M_{*}$ distribution with 
 a width increasing slowly with magnitude.
In contrast, 
 blue galaxies that contribute increasingly at fainter magnitudes 
 have broader $M_{*}$ distributions. 
These broader distributions steepen the low mass end of the SMF
 relative to the slope of the LF. 
 
There are only a few published cluster SMFs extending to 
 $\log (M_{*}/M_{\odot}) \sim 10.0$. 
\citet{Vulcani11} obtain a similar slope,
  $\alpha = -0.987 \pm 0.009$,
 for cumulative SMFs for the low-z WINGS cluster sample \citep{Fasano06}. 
They obtain $\alpha = -0.915 \pm 0.026$
 for the SMFs of the EDisCS sample \citep{White05}. 
\citet{Vulcani13} also report similar slopes from cluster samples 
 at redshift $0.3 < z < 0.8$. 
  
\subsection{Velocity Dispersion Function}\label{vdf}

\begin{figure}
\centering
\includegraphics[scale=0.42]{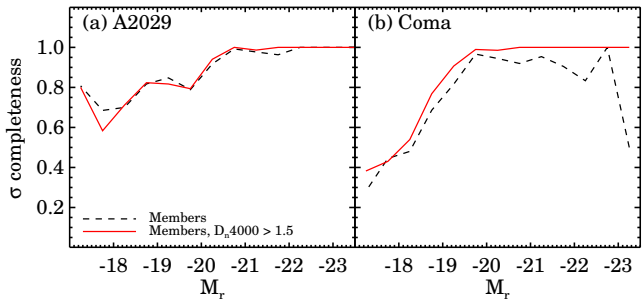}
\caption{
The fractional completeness of $\sigma$ measurements for 
 (a) A2029 members and (b) Coma members. 
The solid lines and the dashed lines represent all members and 
 quiescent members with $\dn > 1.5$, respectively. } 
\label{sigcomp}
\end{figure}

The central velocity dispersion of a galaxy 
 reflects the stellar kinematics. 
The central velocity dispersion is a 
 stellar luminosity weighted sum over objects 
 within the fiber aperture. 
We correct this observed dispersion to a $\sigma$ measured within a 3 kpc radius (see Section 2.1.4).
This dispersion may be proportional to 
 the dispersion of the DM halo and 
 thus it may be a fundamental observable for 
 studying the DM halo distribution (see \citealp{Zahid16c}).
The velocity dispersion is a reasonable halo mass proxy 
 for early-type galaxies dominated by random motions
 \citep{Wake12, vanUitert13, Bogdan15, Zahid16c}. 
For late-type galaxies, 
 the circular velocity is important for characterizing the disk \citep{Sheth03}. 

We examine the velocity dispersion function  
 only for quiescent cluster members with $\dn > 1.5$. 
This criterion conservatively identifies early-type galaxies 
 mainly consisting of an older stellar population
 \citep{Kauffmann03, Woods10, Zahid16c}. 
Previous studies of velocity dispersion functions for field galaxies 
 (e.g. \citealp{Sheth03, Choi07, MonteroDorta16})
 use a variety of definitions for early-type galaxies.
Here we use a homogeneous spectroscopic definition. 
\citet{Moresco13} examine the dependence of 
 the properties of quiescent galaxies in the zCOSMOS-20k spectroscopic sample
 on the quiescent galaxy selection algorithm. 
They show that each classification method yields somewhat different subsamples of galaxies,
 but the overall properties of the quiescent galaxy samples are 
 insensitive to the classification scheme. 

Obtaining a $\sigma$ 
 depends strongly on the signal-to-noise ratio of each spectrum. 
Thus, we lack a $\sigma$ measurement for some cluster members 
 even though we have a reliable redshift.
Furthermore, there are a few $\sigma$ measurements 
 with very large uncertainty. 
Hereafter, we use only $\sigma$s with an error $< 100~\kms$. 
Figure \ref{sigcomp} shows 
 the completeness of $\sigma$ measurements
 for A2029 and Coma members. 
The $\sigma$ completeness for quiescent members
 is much higher than for the entire sample. 
For quiescent galaxies with $\dn > 1.5$, 
 the $\sigma$ measurements are $> 80\%$ complete to $M_{r} < -19$ for A2029 and 
 $> 80\%$ complete to $M_{r} < -19.5$ for Coma.

\begin{figure}
\centering
\includegraphics[scale=0.42]{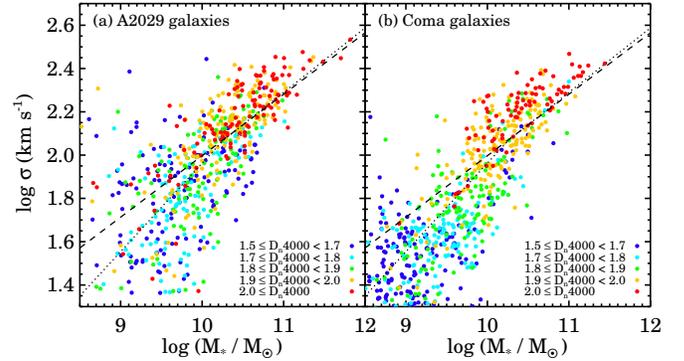}
\caption{
Central velocity dispersion ($\sigma$) vs. stellar mass 
 for (a) A2029 members and (b) Coma members. 
The color displays $\dn$. 
Dashed and dotted lines show the relation 
 derived for SHELS and SDSS \citep{Zahid16c}. }
\label{sigmass}
\end{figure} 
 
Figure \ref{sigmass} displays
 $\sigma$ as a function of $M_{*}$
 for A2029 and Coma members.
For comparison, 
 we plot the relation for field galaxies
 derived by \citet{Zahid16c} 
 using a local SDSS sample ($0.0 < z< 0.2$) and  
 an intermediate redshift ($0 < z < 0.7$) sample from  
 the Smithsonian Hectospec Lensing Survey 
 (SHELS; \citealp{Geller05, Geller14, Geller16}) F2 field.
\citet{Zahid16c} show that for massive galaxies, the slope and zero-point of the 
 $M_{*}$-$\sigma$ relation does not significantly evolve at $z < 0.3$. 
Although the redshift ranges of the two field samples differ,
 the slopes at high $M_{*}$ ($\log (M_{*}/M_{\odot}) > 10.2$) are similar.

The cluster galaxies follow the trend of field galaxies 
 except at the very massive end
 where some outliers in the cluster samples 
 lie above the trend for field galaxies, 
 i.e. these cluster objects have larger $\sigma$ at a given $M_{*}$. 
These outliers with $\dn \gtrsim 2.0$ consist of 
 an older stellar population and may be metal rich. 
Field samples also contain galaxies with $\dn > 2.0$ 
 (Figure 1 in \citealp{Zahid16c}),
 but they are rare.
These large $\dn$ field galaxies 
 also tend to have larger $\sigma$ at a fixed stellar mass
 (Zahid et al. (2017) in preparation). 
The offset between cluster and field galaxies at the massive end 
 reflects differences in the $\dn$ distributions of cluster and field samples.
  
As for the SMFs, 
 we correct the velocity dispersion function for two types of missing members. 
First, we lack $\sigma$ measurements for 
 $\sim 15\%$ of A2029 members and $\sim 18\%$ of Coma members, respectively.
These members have reliable redshifts,
 but the spectrum is not available. 
There are also missing $\sigma$s for probable members 
 resulting from spectroscopic incompleteness. 
  
We derive $\sigma$ for missing members 
 based on conditional probability distributions of $\sigma$
 given the absolute magnitude and $\dn$, 
 $P(\sigma|\dn|M_{r})$. 
As shown in Figure \ref{sigmass}, 
 the conditional probability distribution of $\sigma$ 
 depends strongly on $\dn$. 
We derive these distributions empirically for 
 SDSS galaxies in different $\dn$ ranges. 
We set five bins with $1.5 < \dn \leq 2.0$ with bin size of 0.1, and 
 we reserve one bin for $2.0 < \dn < 3.0$.
The bin sizes are chosen to have similar numbers of galaxies in each bin. 
We also determine the probability distribution  
 for the entire SDSS sample with $\dn > 1.5$. 
We use this probability distribution 
 for cluster members lacking a $\dn$ and a $\sigma$ measurement. 

\begin{figure}
\centering
\includegraphics[scale=0.5]{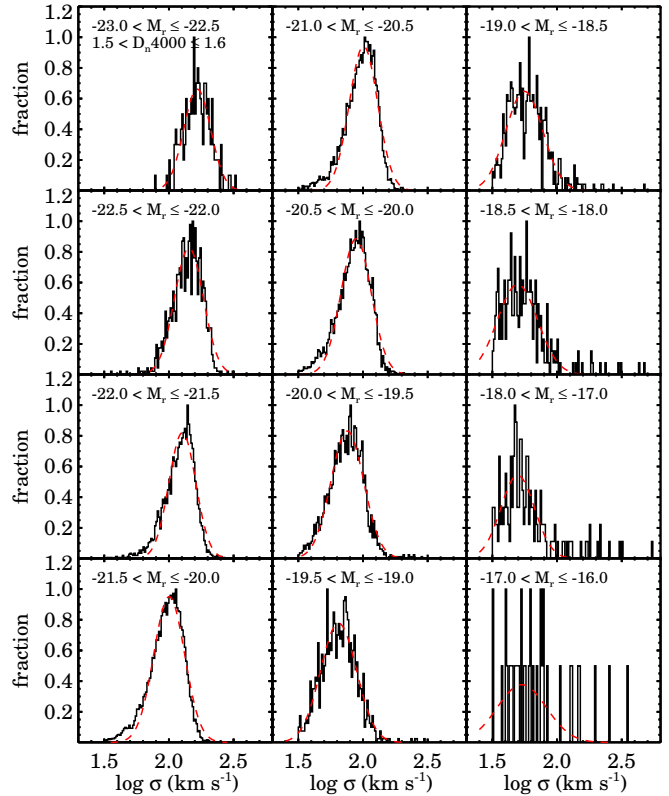}
\caption{
Central velocity dispersion ($\sigma$) distribution for
 SDSS field galaxies with $1.5 < \dn \leq 1.6$
 for various absolute magnitude ranges. 
The dashed lines show Gaussian fits to the $\sigma$ distributions. } 
\label{sigcor}
\end{figure}

Figure \ref{sigcor} shows the 
 conditional probability distributions $P(\sigma | 1.5 < \dn \leq 1.6 | M_{r})$
 for SDSS galaxies in different fixed magnitude ranges and fixed $\dn$ ranges. 
The dashed lines in Figure \ref{sigcor} represent 
 the best-fit Gaussian for the set of $P(\sigma | 1.5 < \dn \leq 1.6 | M_{r})$
 distributions. 
The distributions are skewed to low $\sigma$
 in some magnitude bins. 
However, we ignore the contribution of low $\sigma$ objects 
 because the $\sigma$ measurement for SDSS galaxies are reliable only to 
 $\sigma \sim 60~\kms$ \citep{Thomas13}. 
The distributions for the other $\dn$ bins 
 can also be fit with Gaussians,
 but with different means and widths.
Note that the widths of the conditional probability distributions, 
 $\sim 40~\kms$ at mean $\sigma$ of $100~\kms$, 
 are larger than typical uncertainty in the $\sigma$ measurements
 for both the SDSS and Hectospec samples.
We use the set of best-fit Gaussians 
 to construct a set of $P(\sigma|\dn|M_{r})$
 based on the SDSS field samples. 
 
We compute $\sigma$ for missing members 
 using the set of $P(\sigma|\dn|M_{r})$. 
We know the absolute magnitudes for each member or probable member
 without a $\sigma$ measurement. 
For members missing because of spectroscopic incompleteness, 
 we randomly select a galaxy in the appropriate magnitude bin. 
We calculate $\sigma$ for members and probable members 
 by drawing randomly from the appropriate
 $P(\sigma|\dn|M_{r})$. 
We repeat the process for all missing members 1000 times and 
 measure the velocity dispersion function using each of these 1000 samples. 
Finally, we take the mean of the 1000 velocity dispersion functions 
 as the `corrected' velocity dispersion function of the cluster. 
 
\begin{figure*}
\centering
\includegraphics[scale=0.6]{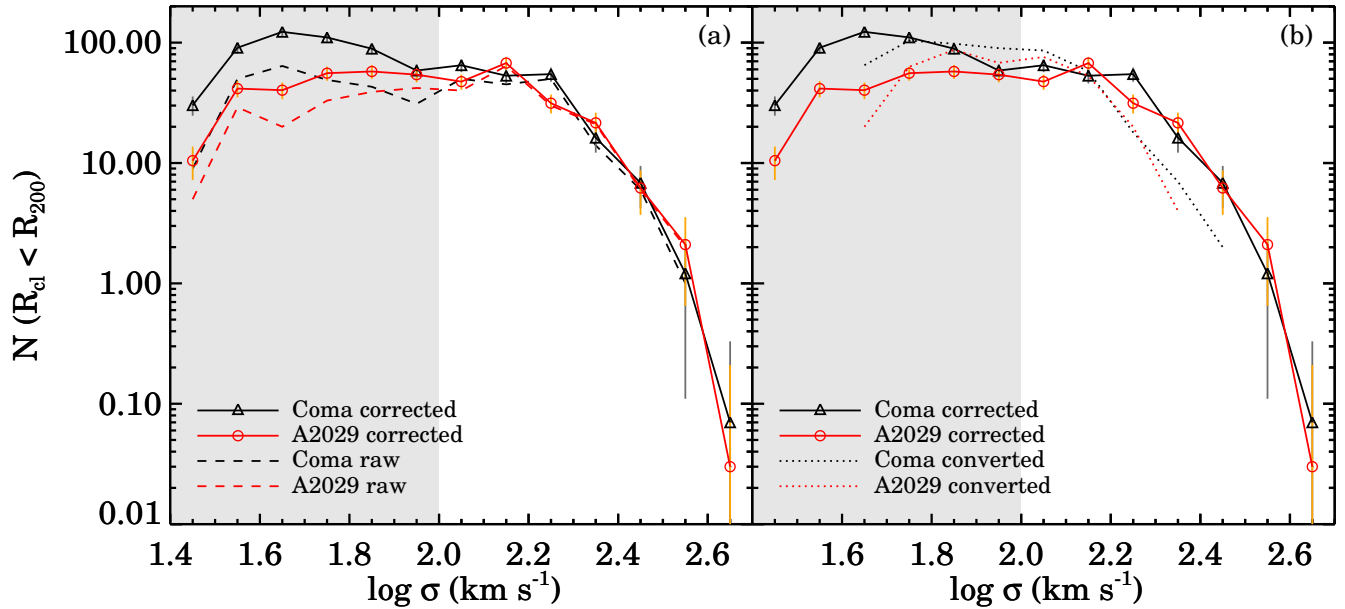}
\caption{
VDF of Coma (triangles) and A2029 (circles)
 compared with (a) the raw VDF (dashed lines) and 
 (b) the VDF assuming the mean relation between $\sigma$ and $M_{r}$ (dotted lines) 
 for each cluster.  
The shaded region indicates the range where incompleteness becomes significant.} 
\label{vf}
\end{figure*}

Figure \ref{vf} shows 
 the corrected velocity dispersion functions (VDFs) 
 for A2029 and Coma. 
The left panel of Figure \ref{vf} compares 
 the VDFs with the raw VDFs. 
The corrections for missing members are negligible 
 for $\log \sigma > 1.9$ and become significant at lower $\sigma$.  

Because the $\sigma$ distribution at each $\dn$ and $M_{r}$ is broad, 
 it is critical to reconstruct the VDF by drawing 
 from the conditional probability distributions \citep{Sheth03}.
To demonstrate this issue, 
 we compare the corrected VDF with a VDF derived by taking the mean $\sigma$ 
 for each absolute magnitude (right panel of Figure \ref{vf}). 
Here we estimate the mean $\sigma$ as a function of absolute magnitude 
 using the SDSS field sample with $\dn > 1.5$. 
We calculate the mock $\sigma$ for all A2029 and Coma members 
 according to their absolute magnitude.  
The converted VDFs differ from the observed VDFs 
 in the sense emphasized by \citet{Sheth03}. 
As \citet{Sheth03} demonstrate, 
 the direct conversion of $\sigma$ from the absolute magnitudes 
 introduces a strong biases in the $\sigma$ distribution. 

As a test of our correction method, 
 we also generate mock cluster VDFs 
 using the set of $P(\sigma|\dn|M_{r})$. 
For these mock VDFs, 
 we draw $\sigma$ for every cluster member. 
These mock VDFs are very similar to the corrected VDFs. 
This test substantiates the correction we apply 
 based on the $P(\sigma|\dn|M_{r})$
 to compensate for missing $\sigma$s. 
 
The VDFs of A2029 and Coma are essentially identical
 for $\log \sigma > 2.0$, 
 where the corrections for incompleteness are negligible.  
The remarkably identical shapes of two cluster VDFs 
 are consistent with the similar shapes of the LFs and SMFs. 
The larger difference between the two VDFs toward low $\sigma$ 
 probably results from the relatively lower spectroscopic incompleteness of A2029.

The similarity of the Coma and A2029 VDFs for $\log \sigma > 2.0$ suggests that 
 the underlying DM subhalo mass distributions of the two target clusters are similar. 
Further studies based on larger cluster samples with 
 different redshift, mass, and dynamical stage 
 may thus provide interesting new probes of the subhalo mass distribution and its evolution. 

%=============================================================
\section{DISCUSSION}

Taking advantage of an intensive spectroscopic survey
 based on SDSS and Hectospec observations, 
 we measure the LFs, SMFs and VDFs for Coma and A2029. 
There are several systematic issues in 
 determining the shape of these functions. 
One critical example is 
 the conditional probability distribution functions we use
 for correcting SMFs and VDFs. 
Because $M_{*}$ and $\sigma$ depend on galaxy properties 
 including colors and $\dn$ (Figure \ref{mfcor} and Figure \ref{sigcor}), 
 a direct translation from absolute magnitude to a mean $M_{*}$ or a mean $\sigma$ 
 introduces significant systematic biases. 
Use of the conditional probability distribution 
 has a more critical impact on the VDF than on the SMF
 because the typical spread in $P(\sigma|\dn|M_{r})$ is much larger than for $P(M_{*}|M_{r})$. 
The differences in the conditional probability distribution 
 result in a substantially different shape for the VDFs relative to the LFs; 
 in contrast the SMFs are similar to LFs. 
The differences in the conditional probability distribution
 result in a substantially different shape for the VDFs relative to the LFs. 
In contrast, the shapes of the SMFs are similar to LFs.  
We examine other systematic effects on the VDF in Section \ref{system}. 
  
Comparing observed SMFs and VDFs with simulations
 tests our understanding of galaxy properties in dense environments.
In Section \ref{comp}, 
 we compare our observed SMFs to the SMFs derived from numerical simulations \citep{Behroozi13, Lim17}.
These simulations provide 
 appropriate model SMFs measured for mock clusters with similar mass to 
 Coma and A2029. 
We compare the observed VDFs for quiescent galaxies only with other observations (Section \ref{imp})
 because current simulations do not compute the VDF in a way that mimics the observations directly.
In particular, 
 to compare directly to observations, 
 model predictions of VDFs need to be made for 
 the luminosity-weighted velocity dispersion within a projected cylinder of fixed aperture. 
Unfortunately, no such model predictions are currently available.

\subsection{Systematic Effects}\label{system}

At face value, 
 the measurement uncertainties of $\sigma$ for Hectospec spectra 
 are larger than for SDSS spectra ($17~\kms$ and $7~\kms$ respectively). 
However, these measurement uncertainties are strongly correlated with the S/N ratio, 
 and thus they are also correlated with apparent magnitude. 
When we compare $\sigma$ measurement errors for A2029 members with $r < 17.77$, 
 the typical uncertainties for SDSS and Hectospec are identical ($\sim 7~\kms$). 
Larger $\sigma$ measurement errors for Hectospec apply to fainter targets. 
These fainter Hectospec targets are mostly galaxies with $\sigma \lesssim 100~\kms$
 where the A2029 VDF becomes incomplete. 
Thus, measuring the cluster VDF based on the data from two different instruments does not affect the 
 shape of VDF at $\sigma > 100~\kms$. 

Despite the systematic differences in the observations, 
 the A2029 VDF is essentially identical to the Coma VDF.  
The aperture correction we apply
 provides a consistent $\sigma$ for A2029 and Coma members. 
Because Coma members are generally brighter than A2029 members, 
 the typical error in $\sigma$ for Coma members ($\sim 4~\kms$) is 
 slightly smaller than for A2029 members.
These differences in $\sigma$ measurement errors for Coma and A2029 members
 have a negligible effect on the shape of VDFs. 

Figure \ref{sigcor} demonstrates that 
 the larger $\sigma$ measurement errors for 
 the faint targets or A2029 members  
 have little impact on determining the shape of VDF. 
The $\sigma$ distribution for quiescent galaxies 
 with a fixed $\dn$ and $M_{r}$ range 
 has an intrinsic dispersion which is significantly larger than 
 the typical $\sigma$ measurement errors for data 
 from both SDSS and Hectospec.
For example, at $M_{r} \sim -20$, 
 the typical measurement uncertainty is 0.05 in $\sigma \sim 100\kms$, 
 while the intrinsic width of the conditional probability distribution is $\sim 0.15$ in $\sigma \sim 100\kms$.
In other words, the intrinsic spread of $\sigma$ for quiescent galaxies is 
 a more important determinant of the shape of VDFs than the measurement error.

\begin{figure}
\centering
\includegraphics[scale=0.5]{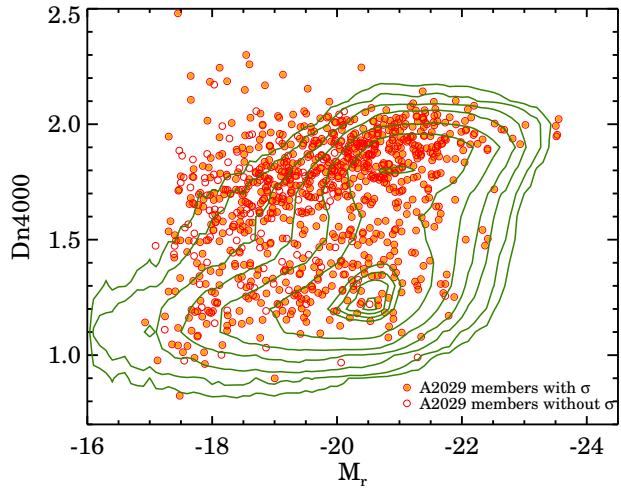}
\caption{$\dn$ vs. $M_{r}$ for A2029 members (symbols)
 compared with the SDSS field sample (contours). 
Filled and open circles show A2029 members with $\sigma$ and without $\sigma$, respectively.}
\label{dn}
\end{figure}
 
To correct for missing $\sigma$s for cluster members, 
 we derive the $P(\sigma|\dn|M_{r})$ using the SDSS field sample.
The $\dn$ distribution for the field actually differs somewhat from the one for cluster members. 
Figure \ref{dn} shows $\dn$ versus $M_{r}$ for A2029 members 
 compared with the SDSS field sample (contour). 
The SDSS field sample lacks galaxies with $M_{r} < -19$ and $\dn > 1.8$;  
 most cluster members without $\sigma$ measurements appear in this domain. 
However, the lack of appropriate $P(\sigma|\dn|M_{r})$ for 
 these missing members has little effect on the shape of VDFs. 
Missing cluster members with $M_{r} < -19$ and $\dn > 1.8$
 tend to have $\sigma < 100~\kms$.
Thus the correction is insignificant for $\sigma > 100~\kms$ 
 (left panel of Figure \ref{vf}), 
 where we examine the shape of the cluster VDFs. 
 
\subsection{Comparison of the SMFs with Simulations}\label{comp}

An accurate determination of the LF, SMF, and VDF
 provides important constraints on galaxy formation models. 
Traditionally, 
 the LF has been compared to the DM subhalo mass distribution. 
\citet{Ferrarese16} show that 
 the Virgo LF (and also SMF) is significantly shallower than 
 the expected distribution from $\Lambda$CDM 
 ($\alpha \sim -1.9$, e.g. \citealp{Springel08}). 
Other spectroscopic LFs including A85 \citep{Agulli16} and A2199 \citep{Rines08}
 are also less steep than the subhalo mass distributions derived from simulations. 
However, 
 the complexity of the transformation from subhalo mass to luminosity is non-trivial
 and thus these comparisons are hard to interpret.

SMFs provide a somewhat more direct basis for comparison with the models. 
Unlike the LF,
 the SMF attempts to correct for variations in the stellar populations of galaxies. 
\citet{Lim17} compare empirical models of SMFs 
 with observed SMFs in groups/clusters of different halo masses \citep{Lan16}. 
For massive clusters with halo mass $\log (M_{\rm halo} / M_{\odot}) > 13.7$, 
 the empirical models are consistent with the observed SMFs. 
They also compare the SMFs obtained from hydrodynamic simulations, 
 i.e. Illustris \citep{Vogelsberger14} and EAGLE \citep{Schaye15};
 these simulations do not appear to agree as well with the observed SMFs 
 (see Figure 5 of \citealp{Lim17}). 

\begin{figure}
\centering
\includegraphics[scale=0.5]{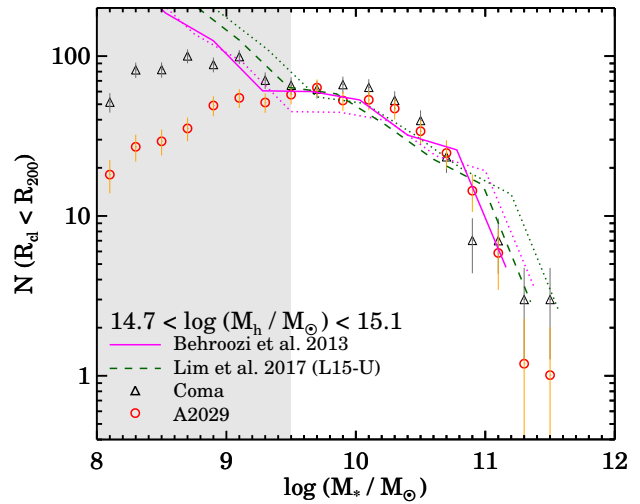}
\caption{
Observed stellar mass functions for Coma (triangles) and A2029 (circles)
 compared with the model stellar mass functions 
 from \citet{Behroozi13} (magenta solid lines) and from \citet{Lim17} (green dashed lines). 
The model stellar mass functions are scaled 
 to compare the overall shape to the observed stellar mass functions. 
The magenta dotted lines and green dotted lines 
 show the original unscaled model stellar mass functions
 from \citet{Behroozi13} and \citet{Lim17}, respectively. }
\label{mfsim}
\end{figure}

In Figure \ref{mfsim}, we compare the Coma and A2029 SMFs with
 the empirical model SMFs from \citet{Lim17}. 
For simplicity, 
 we employ two empirical model SMFs from \citet{Behroozi13} and \citet{Lim17} 
 (both kindly provided by S.H.Lim)
 for a halo mass in the range $14.7 < \log (M_{\rm halo} / M_{\odot}) < 15.1$ 
 corresponding to the Coma and A2029 dynamical masses (i.e. $M_{200}$).
This range is the most massive range sampled by the models and 
 is most appropriate for comparison with Coma and A2029. 
To account for any systematic difference in the $M_{*}$ and the amplitude, 
 we scale the models to match the observed cluster SMFs
 using $\chi^{2}$ minimization. 
For completeness,  
 we show both the scaled and the unscaled model SMFs in Figure \ref{mfsim}. 
 
The overall shapes of the model SMFs, regardless of scaling, 
 match the observed SMFs 
 for $\log (M_{*}/M_{\odot}) > 9.5$ remarkably well.
Interestingly, 
 both models account for the flat portion of the observed SMFs
 ($9.5 < \log (M_{*}/M_{\odot}) < 10.5$). 
The observed cluster SMFs are incomplete for $\log (M_{*}/M_{\odot}) < 9.5$; 
 thus we cannot test the upturn that appears in both models. 
At the massive end, $\log (M_{*}/M_{\odot}) > 11.0$, 
 the scaled model SMFs predict too few massive galaxies. 
This difference is in the same direction that 
 \citet{Munari16} find between model and observed VDFs for $\log \sigma > 2.0$. 
However, the uncertainty of the model SMFs 
 at $\log (M_{*}/M_{\odot}) > 11.0$ is large \citep{Lim17}.  
Nonetheless, the subtle discrepancy could ultimately be a test of 
 models for the formation of the most massive galaxies in clusters. 
 
\subsection{Implications of the Cluster VDFs}\label{imp}

The statistical study of the VDF for quiescent galaxies in clusters
 complements previous studies of the VDF based on field samples. 
Galaxy clusters are a useful testbed 
 because the cluster galaxies are essentially at a fixed distance and
 share the same dense environment. 
VDFs derived from cluster samples 
 may differ from VDFs based on well controlled field samples
 as a result of density dependent processes affecting galaxy evolution. 

\begin{figure}
\centering
\includegraphics[scale=0.5]{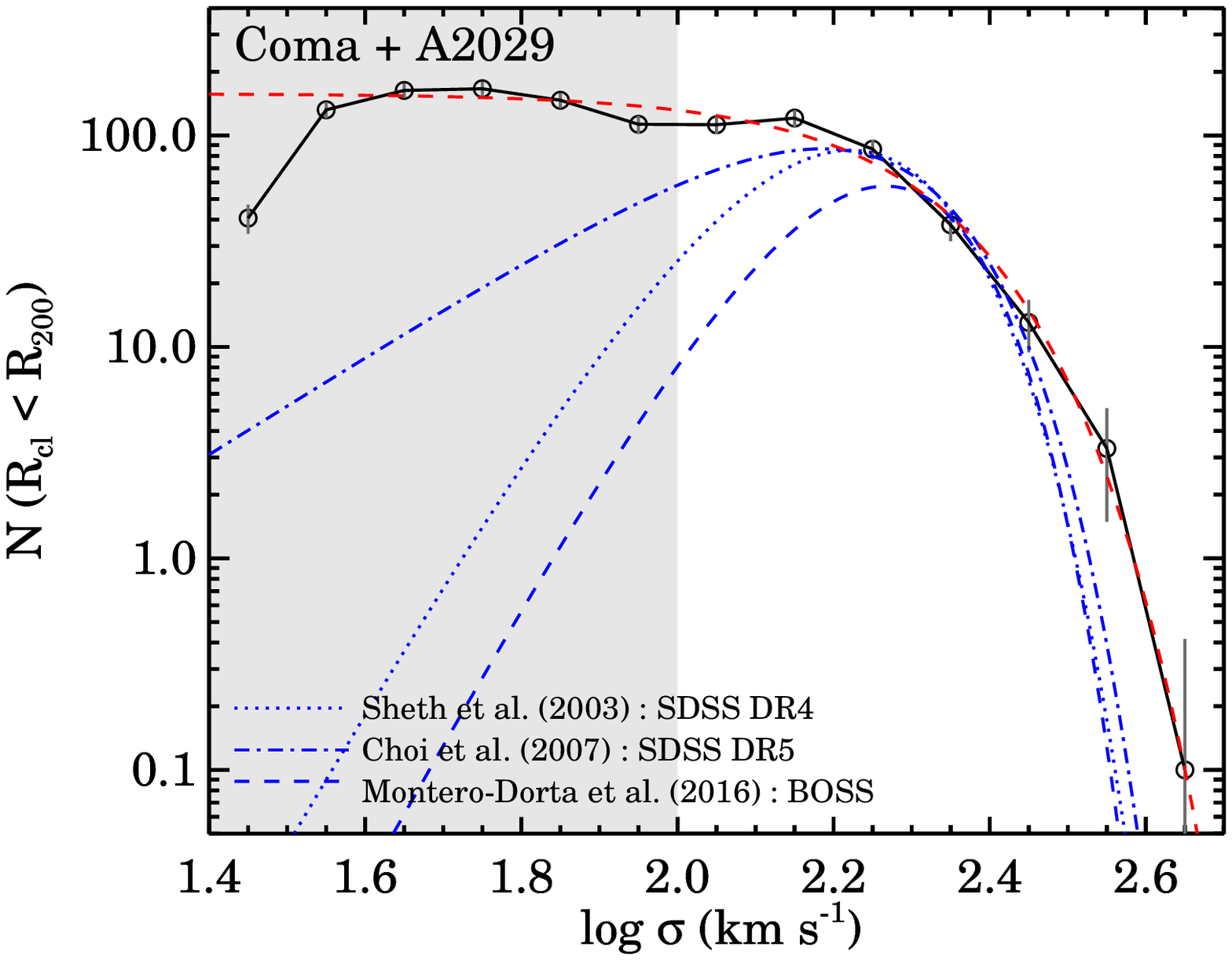}
\caption{
Combined velocity dispersion function (VDF) for A2029 and Coma
 (solid lines with data points). 
Red dashed line displays 
 the best-fit modified Schechter function for the combined cluster VDFs. 
Blue dotted, dot-dashed and dashed lines show 
 fitting functions for the velocity dispersion functions derived from  
 SDSS field galaxies \citep{Sheth03, Choi07} and 
 for BOSS field galaxies \citep{MonteroDorta16}, respectively. 
Note the differences in both the high and low $\sigma$ range. } 
\label{vfsum}
\end{figure}

Figure \ref{vfsum} displays 
 the combined cluster VDF for quiescent cluster members with $\dn > 1.5$
 (the sum of the Coma and A2029 VDFs). 
For comparison, 
 we also plot the VDFs 
 from the SDSS field samples \citep{Sheth03, Choi07} and 
 from the BOSS field sample \citep{MonteroDorta16}. 
The field VDFs are described by a `modified' Schechter function, 
\begin{equation}
\phi(V)dV = \phi_{0}~10^{\alpha(V-V_{*})}~exp[-10^{\beta(V-V_{*})}]\frac{\beta~ln10}{\Gamma(\alpha/\beta)}~dV,
\end{equation}
 where V is $\log \sigma$. 
The field VDFs in Figure \ref{vfsum}
 show the best-fit modified Schechter functions taken from the respective references. 
We also fit the combined cluster VDF with the modified Schechter functions
 at $\log \sigma > 1.6$. 
Because the combined cluster VDF is much flatter than the field VDFs at lower $\sigma$, 
 the best-fit parameters differ from the field VDFs. 
Table \ref{vftab} lists the best-fit parameters for the cluster VDF 
 and those from the literature. 

\begin{deluxetable*}{lcccc}
\tablecolumns{5}
\tabletypesize{\scriptsize}
\tablewidth{0pt}
\setlength{\tabcolsep}{0.05in}
\tablecaption{Velocity Dispersion Function Parameters}
\tablehead{
\colhead{Source} & \colhead{$\sigma$ range} & \colhead{$\alpha$} & \colhead{$\beta$} & \colhead{$\sigma_{*}$} }
\startdata
This study             & $\log \sigma > 1.6$  & $0.00 \pm 0.23$ & $2.47 \pm 0.63$ & $194.85 \pm 26.46$ \\
\citet{Sheth03}        & $\log \sigma > 1.95$ & $6.5$           & $1.93$          & $88.8$             \\
\citet{Choi07}         & $\log \sigma > 1.84$ & $2.32 \pm 0.10$ & $2.67 \pm 0.07$ & $161.00 \pm  0.05$ \\
\citet{Chae10}         & $\log \sigma > 1.9$  & $0.85$          & $3.27$          & $217.0$            \\
\citet{MonteroDorta16} & $\log \sigma > 2.35$ & $6.75 \pm 0.99$ & $2.37 \pm 0.14$ & $118.86 \pm 12.40$
\enddata
\label{vftab}
\end{deluxetable*} 
 
To compare the overall shape of the VDFs, 
 we scale the field VDFs to match the amplitude of the combined cluster VDF. 
Unlike the cluster VDF, 
 the field VDFs are normalized by the survey volume. 
Thus, we scale the amplitude to compare the cluster and field VDFs. 
Both field VDFs and the cluster VDF are based on $\sigma$ from SDSS/BOSS and 
 comparable Hectospec data. 
Although the aperture correction methods differ among these studies, 
 we sample for the $\sigma$ bins for field VDFs without additional calibration to the $\sigma$ aperture we use. 
The typical aperture correction is only a few \% \citep{MonteroDorta16}. 
Thus, this difference is negligible for the comparison we make here.

Figure \ref{vfsum} underscores the differences 
 between the cluster and field VDFs. 
Although we scale the field VDFs, 
 the shape difference is significant. 
At high $\sigma$ ($\log \sigma > 2.4$), 
 the combined cluster VDF substantially exceeds any field VDF. 
The large $\sigma$ galaxies appearing in clusters are 
 the brightest cluster galaxies (BCGs),
 which are rare in field samples.
 
The discrepancy between cluster and field VDFs is even larger at low $\sigma$; 
 the field VDFs decline rapidly with respect to the cluster VDF. 
Because the BOSS VDF is limited to $\log \sigma > 2.35$, 
 the difference may not be surprising.
The SDSS VDFs are substantially shallower than the cluster VDF 
 although they are complete to $\log \sigma > 2.0$.
 
Different early-type galaxy selection schemes 
 may result in the VDF difference toward low $\sigma$, 
 but not high $\sigma$ \citep{Choi07}. 
Our $\dn$ selection differs from previous approaches.
However, the shape of the combined cluster VDF 
 appears to be insensitive to the specific classification method. 
For example, the shape of the combined cluster VDF is the same
 when we measure the cluster VDF based on red-sequence member galaxies 
 rather than galaxies with $\dn > 1.5$.
We note that morphological classification based on the appearance in SDSS images 
 is inadequate for galaxies at the redshift of A2029. 

The A2029 and Coma VDFs represent a lower limit to the low $\sigma$ cluster VDF. 
Missing faint galaxies with $M_{r} < -20$ 
 tend to have generally low $\sigma$.
There are also low surface brightness galaxies 
 with $\mu_{r} < 24$ mag arcsec$^{-2}$
 missing from our sample. 
Thus, the cluster VDF would be steeper at the low $\sigma$ end
 if it were more complete. 

Although the cluster VDFs represent a lower limit at low $\sigma$, 
 the observed cluster VDFs are already much steeper than the field VDFs.
The discrepancy between these clusters and the field suggests that 
 corrections made to the field VDFs to account for missing low $\sigma$ galaxies 
 may be inadequate.
\citet{Choi07} measure the SDSS field VDF 
 for galaxies with $M_{r} < -16.8$ in several volume-limited samples.
However, a sample that is volume-limited 
 is not equivalent to a velocity dispersion-limited sample
 (Zahid et al. 2017, in preparation). 
Making the necessary correction for this difference is challenging. 

Differences between cluster and field VDFs 
 could be an important window for the related halo mass distribution.
The origin of the differences toward high $\sigma$ is an interesting issue
 because high $\sigma$ BCGs appear predominantly in clusters. 
In order to interpret the differences between the cluster and field VDFs
 from a deeper astrophysical perspective, 
 samples that are homogeneous  
 in early-type classification, spectroscopic completeness,
 and statistics are required both for clusters and the field. 

%=============================================================
\section{SUMMARY}
 
We use dense redshift surveys from SDSS and MMT/Hectospec
 to identify spectroscopic members of two massive clusters, Coma and A2029. 
We identify essentially complete samples of $\sim 1000$ spectroscopic members for each cluster 
 based on the caustic technique. 
To date, only Coma and A2029 have such large samples of spectroscopically identified members. 
Using the spectroscopic members, 
 we measure the luminosity functions (LFs), 
 stellar mass functions (SMFs), and 
 velocity dispersion functions (VDFs) for these systems. 

The bright end of the cluster LFs is identical
 to the other cluster LFs derived based on spectroscopic membership.
The cluster LFs at the bright end ($M_{r} < -18$) are dominated by quiescent (red sequence) galaxies 
 and the slope tends to be flatter than the LF measured over broader luminosity ranges.  

The cluster SMFs mimic the cluster LFs.
The SMFs are flat to $\log (M_{*}/M_{\odot}) \sim 9.5$
 where the spectroscopic survey is complete. 
However, the SMFs are somewhat steeper than the LFs over the comparable range.
In accounting for missing observations, the translation from luminosity to $M_{*}$ requires use of 
 conditional probability distribution functions.
The resulting observed cluster SMFs are remarkably consistent with 
 simulated SMFs \citep{Behroozi13, Lim17}. 
A subtle difference at $\log (M_{*}/M_{\odot}) \sim 11.0$ is interesting 
 because it suggests that 
 the number of massive halos produced in the simulations 
 may be insufficient to match the observations. 

For the first time, 
 we derive the cluster VDFs for quiescent cluster members 
 over the broad range $\log \sigma > 1.5$. 
The A2029 VDF and Coma VDF are essentially identical.  
This similarity suggests that DM subhalo distributions for these two massive clusters are essentially identical.  

The cluster VDFs differ from published field VDFs 
 at both high and low $\sigma$. 
The cluster VDFs exceed the field VDFs at  $\sigma \gtrsim 250~\kms$
 probably reflecting the presence of massive BCGs in the cluster environment. 
The cluster VDFs also substantially exceed 
 the field VDFs at $\sigma \lesssim 100~\kms$
 despite the fact that the cluster VDFs represent a lower limit to the count of objects at these dispersions. 
The differences between cluster and field VDFs are 
 a promising basis for understanding the velocity dispersion and 
 related halo mass distributions in different environments and at various redshifts.
 
VDFs may be a particularly direct probe of galaxy evolution 
 because several studies suggest that 
 $\sigma$ is a good proxy for the DM subhalo mass (\citealp{Wake12, Bogdan15, Zahid16c}).
Comparison between the observed VDFs and 
 the simulated quantities calculated directly from hydrodynamic simulations 
 (e.g. Illustris;\citealp{Vogelsberger14} or EAGLE;\citealp{Schaye15})
 are crucial for a clearer physical understanding of the astrophysical implications of this measure.  
Like the SMF, the VDF may vary with environment and redshift. 
Combining simulations that properly mimic the observations with more extensive data
 could thus provide a new probe of the formation and coevolution of galaxies and their massive halos. 

%\clearpage

\acknowledgments
We thank Perry Berlind and Michael Calkins for operating Hectospec 
 and Susan Tokarz for helping the data reduction.
This paper uses data products produced by the OIR Telescope Data Center, 
 supported by the Smithsonian Astrophysical Observatory. 
We also thank Charles Alcock and Ian Dell'Antonio for insightful discussions. 
J.S. gratefully acknowledges the support of the CfA Fellowship. 
The Smithsonian Institution supported research of M.J.G. and D.F.
H.J.Z is supported by the Clay Postdoctoral Fellowship. 
A.D. acknowledges support from the grant 
 Progetti di Ateneo/CSP TO Call2 2012 0011 
 “Marco Polo” of the University of Torino, 
 the INFN grant InDark, 
 the grant PRIN 2012 “Fisica Astroparticellare Teorica” of 
 the Italian Ministry of University and Research. 
K.R. is partially supported by a Cottrell College Science Award from the Research Corporation.
This research has made use of NASA’s Astrophysics Data System Bibliographic Services.
Funding for SDSS-III has been provided by the Alfred P. Sloan Foundation, 
 the Participating Institutions, the National Science Foundation, and 
 the U.S. Department of Energy Office of Science. 
The SDSS-III web site is http://www.sdss3.org/.
SDSS-III is managed by the Astrophysical Research Consortium for 
 the Participating Institutions of the SDSS-III Collaboration including the University of Arizona, 
 the Brazilian Participation Group, Brookhaven National Labora- tory, 
 University of Cambridge, Carnegie Mellon University, University of Florida, 
 the French Participation Group, the German Participation Group, Harvard University, 
 the Instituto de Astrofisica de Canarias, 
 the Michigan State/Notre Dame/ JINA Participation Group, 
 Johns Hopkins University, Lawr- ence Berkeley National Laboratory, 
 Max Planck Institute for Astrophysics, Max Planck Institute for Extraterrestrial Physics,
 New Mexico State University, New York University, Ohio State University, 
 Pennsylvania State University, University of Portsmouth, Princeton University, 
 the Spanish Participation Group, University of Tokyo, University of Utah, 
 Vanderbilt University, University of Virginia, University of Washington, and Yale University.
\facility{MMT}

%=============================================================================================================
%  Bibliograph - References
%=============================================================================================================
\bibliographystyle{apj}
\bibliography{ms}

\clearpage

\end{document}